\documentclass{ws-ijmpa}
\usepackage[super,compress]{cite}

\newcommand{\be}{\begin{equation}}
\newcommand{\ee}{\end{equation}}
\newcommand{\bea}{\begin{eqnarray}}
\newcommand{\eea}{\end{eqnarray}}
\newcommand{\beq}{\begin{equation}}
\newcommand{\eeq}{\end{equation}}
\newcommand{\nn}{\nonumber}

\def\ga{\mathrel{\mathpalette\fun >}}
\def\fun#1#2{\lower3.6pt\vbox{\baselineskip0pt\lineskip.9pt
\ialign{$\mathsurround=0pt#1\hfil##\hfil$\crcr#2\crcr\sim\crcr}}}

\begin{document}

\title{ASYMPTOTIC REGIMES FOR HADRON DIFFRACTIVE SCATTERINGS
 AND COULOMB INTERACTION. \\
 ARGUMENTS FOR THE BLACK DISK MODE}

\author{V.V. ANISOVICH, V.A. NIKONOV}
\address{National Research Centre "Kurchatov Institute",
Petersburg Nuclear Physics Institute, Gatchina 188300, Russia}

\author{J. NYIRI}
\address{Institute for Particle and Nuclear Physics, Wigner RCP,
Budapest 1121, Hungary }

\maketitle

\begin{history}
\received{Day Month Year}
\revised{Day Month Year}
\end{history}

\begin{abstract}
Comparative analysis of the interplay of hadron and Coulomb
interactions in $pp^\pm$ scattering amplitudes is performed in a broad
energy interval, $\sqrt{s}=1-10^6$ TeV, for two extreme cases: for the
asymptotic interactions of hadrons in black disk and resonant disk
modes. The interactions are discussed in terms of the $K$-matrix
function technique. In the asymptotic regime the real part of the
hadronic amplitude is concentrated in both cases on the boundary of the
disks in the impact parameter space but the LHC energy region is not
asymptotic for the resonant disk mode that lead to a specific interplay
of hadronic and coulombic amplitudes. For the $pp$ scattering at
$\sqrt{s}\sim 10$ TeV an interplay of the hadron and Coulomb
interactions in the resonant disk modes is realized in a shoulder in
$d\sigma_{el}/d{\bf q}^2$ at ${\bf q}^2\sim 0.0025-0.0075$ GeV$^2$. The
absence of such a shoulder in the data at 8 TeV can be considered as an
argument against the resonant disk mode.
 \end{abstract}

\ccode{PACS numbers: 13.85.Lg, 13.85.-t, 13.75.Cs, 14.20.Dh}
%\maketitle

\section{Introduction}
New data \cite{8TeV} for $d\sigma_{el}/d|t|$ at $\sqrt{s}=8$ TeV in the region
$|t|=0.0006-0.2061$ GeV$^2$ shed light on the problem of the asymptotic regime
of hadron cross sections at ultrahigh energies.

The analysis of the pre-LHC
\cite{Arnison:1983mm,Bozzo:1984ri,Amos:1990fw,Augier:1993sz,Abe:1993xx}
and LHC
\cite{Latino:2013ued,Aad:2012pw,Khachatryan:2015gka,Abelev:2012sea}
data for hadronic total cross sections and diffractive scattering cross
sections points to a steady growth of the optical density in the impact
parameter space, $T(b,\ln s)$, with increasing energy, $\sqrt s$. At
LHC energy the profile function of the $pp$-scattering amplitude
reaches the black disk limit at small $b$. Two extreme scenarios are
possible at larger energies, $\sqrt{s}\ga 100$ TeV.

First, the profile function gets frozen in the black disk limit,
$T(b)\simeq 1$, while the radius of the black disk, $R_{black\;disk}$,
is increasing with $\sqrt s$ providing $\sigma_{tot}\sim \ln^2s$,
$\sigma_{el}\sim \ln^2s$, $\sigma_{inel}\sim \ln^2s$. The black disk
regime was studied in a number of papers, see refs.
\cite{GW,Dakhno:1999fp,block,ann2013asym} and references therein.

In another scenario the profile function continues to grow at
$\sqrt{s}\ga 100$ TeV approaching the maximal value, $T(b)\simeq 2$,
that means the resonant disk mode. In this mode the disk radius,
$R_{resonant\;disk}\,$, increases providing the $\ln^2 s$-growth of the
total and elastic cross sections, $\sigma_{tot}\sim \ln^2s$,
$\sigma_{el}\sim \ln^2s$, but a slower increase of inelastic cross
section, $\sigma_{inel}\sim \ln s$, see refs.
\cite{Troshin2014,Dremin2014,ann2014reso} .

Hadron physics at ultrahigh energies is a physics of large energy
logarithms, $\ln s\equiv \xi>>1$
\cite{Gaisser1988,Block1990,Fletcher1992} , and increasing parton disks
\cite{as1978,alr1979,ann2013DN,ann2014review,a2015ufn} . The black disk
picture corresponds to the non-coherent parton interactions in hadron
collisions. In the resonant disk mode the partons interact coherently
thus providing a maximal cross section corresponding to the Froissart
limit \cite{Froissart1961} .

In this paper, considering an interplay of hadron and Coulomb
interactions, we concentrate our attention on the resonant disk mode;
the corresponding consideration of the interplay in the black disk mode
was performed in refs.
\cite{ann2015coul-arxiv,ann2015real,Anisovich:2016nss} .

The asymptotic regime in the resonant disk mode starts at essentially larger energies
than in the black disk regime. In the LHC energy region the resonant disk  profile
function increases more rapidly than that in the black disk mode.
It results in a
larger real part of the scattering amplitude and,
correspondingly, in a larger interference of hadronic and coulombic terms.
At $\sqrt{s}\sim 10$ TeV the interplay of hadronic and coulombic interactions reveals
itself in a shoulder
 in $d\sigma_{el}/d{\bf q}^2$ at ${\bf q}^2\sim 0.0025-0.0075$ GeV$^2$.
 The shoulder is not seen in the data for $pp$ scattering at 8 TeV \cite{8TeV}
that provides an argument in favour of the black disk mode.

In Section 2 we calculate the real part of the hadronic amplitude in
the resonant mode. Inclusion of the Coulomb interaction is
performed in Section 3, a comparative analysis of results in the black
disk and resonant disk modes is presented.
In these section we demonstrate the data at 8 TeV \cite{8TeV} versus
calculations of $d\sigma_{el}/d|t|$ in the black disk and resonant disk modes -
the comparison definitely argues
in favour of the black disk mode.

\section{Hadronic amplitude and K-matrix function  }

For the
hadronic scattering amplitude with switched off Coulomb interaction
we write:
\bea \label{1}
A^{H}({\bf q}^2 ,\xi)&=& \int d^2b e^{i{\bf q}{\bf b}}T^{H}(b,\xi),
\nn \\
T^{H}(b,\xi)&=&1-\eta(b,\xi)\exp{(2i\delta(b,\xi)})
= \frac{-2i K^{H}(b,\xi)}{1-iK^{H}(b,\xi)}\ ,
\eea
where $\xi=\ln s$ and $b=|{\bf b}|$. The complex function
$K^{H}(b,\xi)$ presents part of amplitude without elastic rescatterings (or, with
inelastic processes in the intermediate states only).

For the imaginary and real parts of the introduced functions we write
\cite{ann2015coul-arxiv,ann2015real} :
\bea \label{e2}
-iK^{H}(b,\xi)&=&
K^{H}_{\Im}(b,\xi)-iK^{H}_{\Re}(b,\xi)
\simeq K^{H}_{\Im}(b,\xi)-i\frac{\pi}{2}\frac{\partial K^{H}_{\Im}(b,\xi
)}{\partial \xi} \ .
\eea
At the asymptotic regime the imaginary part of the amplitude is a
generating function for
the real part of the amplitude,
the
asymptotic equality $[\sigma_{tot}(pp)/\sigma_{tot}(p\bar
p)]_{\sqrt{s}\to\infty}=1$ is supposed for that, see
refs. \cite{ann2015coul-arxiv,ann2015real} for details. The total and
diffractive cross sections read:
\be
\sigma_{tot}=2 \int d^2b T^{H}_{\Im}(b,\xi),  \qquad
4\pi\frac{d\sigma_{el}}{d{\bf q}^2}=
(1+\rho^2)\Big|A_{\Im}^{H}({\bf q}^2)\Big|^2\,,
\ee
with the usual
notation $A^{H}_{\Re}({\bf q}^2,\xi)/A^{H}_{\Im}({\bf q}^2,\xi)=\rho({\bf q}^2,\xi)$.
Taking into account that $\rho^2$ is small ($\rho^2\sim 0.01$ and
asymptotically $\rho \sim 1/\ln s$) one can
approximate:
\be \label{c5}
\Big|A^{H}_{\Im}({\bf q}^2 ,\xi)\Big| \simeq
2\pi^\frac12\sqrt{
\frac{d\sigma_{el}}{d{\bf q}^2}
},
\ee
Eq. (\ref{c5})
makes possible direct calculations of the real part of the scattering
amplitude, $A^{H}_{\Re}({\bf q}^2 ,\xi)$, on the basis of the energy
dependence of the diffractive scattering cross section. The
corresponding calculations were performed in ref. \cite{ann2015real}
using preLHC
\cite{Arnison:1983mm,Bozzo:1984ri,Amos:1990fw,Augier:1993sz,Abe:1993xx}
and LHC
\cite{Latino:2013ued,Aad:2012pw,Khachatryan:2015gka,Abelev:2012sea}
data. The results were extrapolated in the region $\sqrt{s}=10-10^6$
TeV using the black disk mode hypothesis for the asymptotic regime. The
next step, an inclusion of the Coulomb interaction into consideration
of the scattering amplitude, was made in refs.
\cite{ann2015coul-arxiv,Anisovich:2016nss} .

\subsection{Resonant disk and K-matrix function  }

From the data it follows that both $T^{H}(b)$ and $-i K^{H}(b)$ are
increasing with energy, being less than unity. If the eikonal mechanism
does not quench the growth, both characteristics cross the black disk
limit getting $T^{H}(b)>1$, $-i K^{H}(b)>1$. If $-i K^{H}(b)\to \infty$
at $\ln s\to \infty$, which corresponds to a growth caused by the
supercritical pomeron ($\Delta>0$), the diffractive scattering process
gets to the resonant disk mode.

For following the resonant disk switch-on we use the two-pomeron model
with parameters providing the description of data at 1.8 TeV and 7 TeV
\cite{ann2014reso,ann2016arxiv} , namely:
 \bea
\label{rd5a}
 K_\Im^{H}(b)&=&\int \frac{d^2q}{(2\pi)^2}
\exp{\Big(-i{\bf q}{\bf b}\Big)}
\sum g^2 s^\Delta
e^{-(a+\alpha\xi){\bf q}^2)}  \nn
\\
&=&
   \sum
\frac{g^2}{4\pi(a+\alpha'\xi)}
\exp{\Big[\Delta\xi-\frac{{\bf b}^2}{4(a+\alpha'\xi)}\Big]}\,,
\qquad \xi=\ln \frac {s}{s_0}.
\eea
The real part of the $K$-matrix function is determined in accord with
Eq. (\ref{e2}).

The following parameters are used
for the leading and the next-to-leading pomerons:
\be
\begin{tabular}{l|l|l}
%\hline
    parameters       & leading pole  & next-to-leading    \\
\hline
$\Delta$                    &  0.20          &  0      \\
$\alpha'_P$ [GeV$^{-2}$] &  0.18           &  0.14   \\
$a$   [GeV$^{-2}$]     &  6.67           & 2.22   \\
$g^2$  [ mb ]        &  1.65           &  27.2  \\
$s_0$  [GeV$^{2}$]   &  1               &   1   \\
%\hline
\end{tabular}
\ee
The description of the data at 1.8 TeV and 7 TeV
within the resonant disk mode and neglecting the real
part of the amplitude was performed in ref. \cite{ann2014reso} . The inclusion of
the real part into consideration of these data leads to some corrections of
the parameters, see \cite{ann2016arxiv} . We use here parameters of ref.
\cite{ann2016arxiv} for the presentation of diffractive scatterings at 8 TeV and 14 TeV.

\begin{figure*}[ht]
%\Fig. 1
\centerline{\epsfig{file=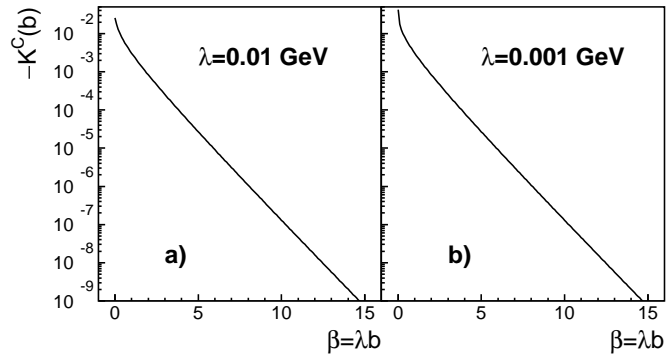,width=0.70\textwidth}}
\caption{The $K$-matrix function, $-K^C(b)$, for the pure Coulomb interaction
in $pp$ collision at different $\lambda$, we use $\beta= \lambda b$ for
abscissa with a) $\lambda =0.01$ GeV and b) $\lambda =0.001$ GeV.}
\label{fig1}
\end{figure*}

\begin{figure*}[ht]
%\Fig. 2
\centerline{\epsfig{file=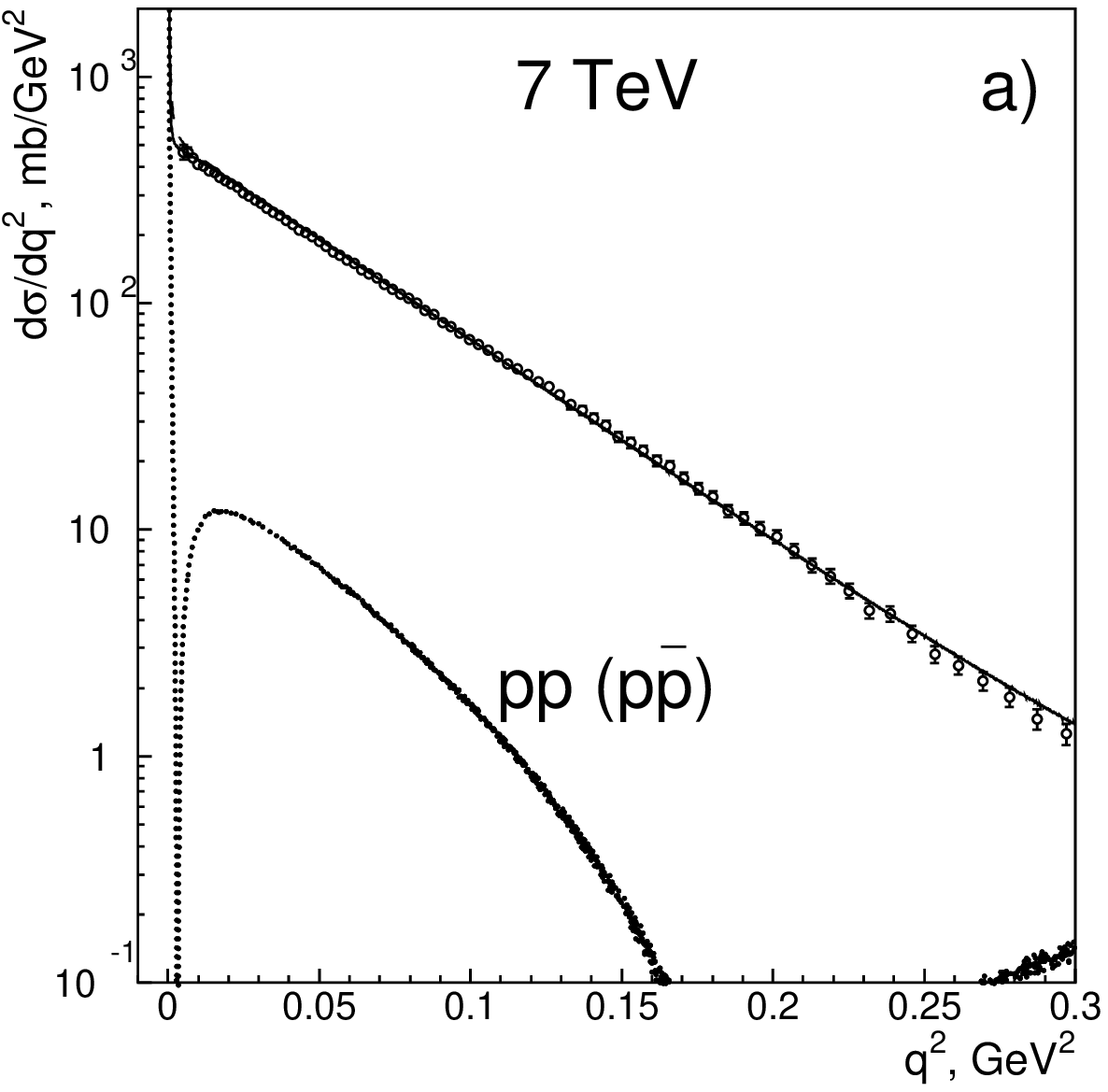,width=0.45\textwidth}
            \epsfig{file=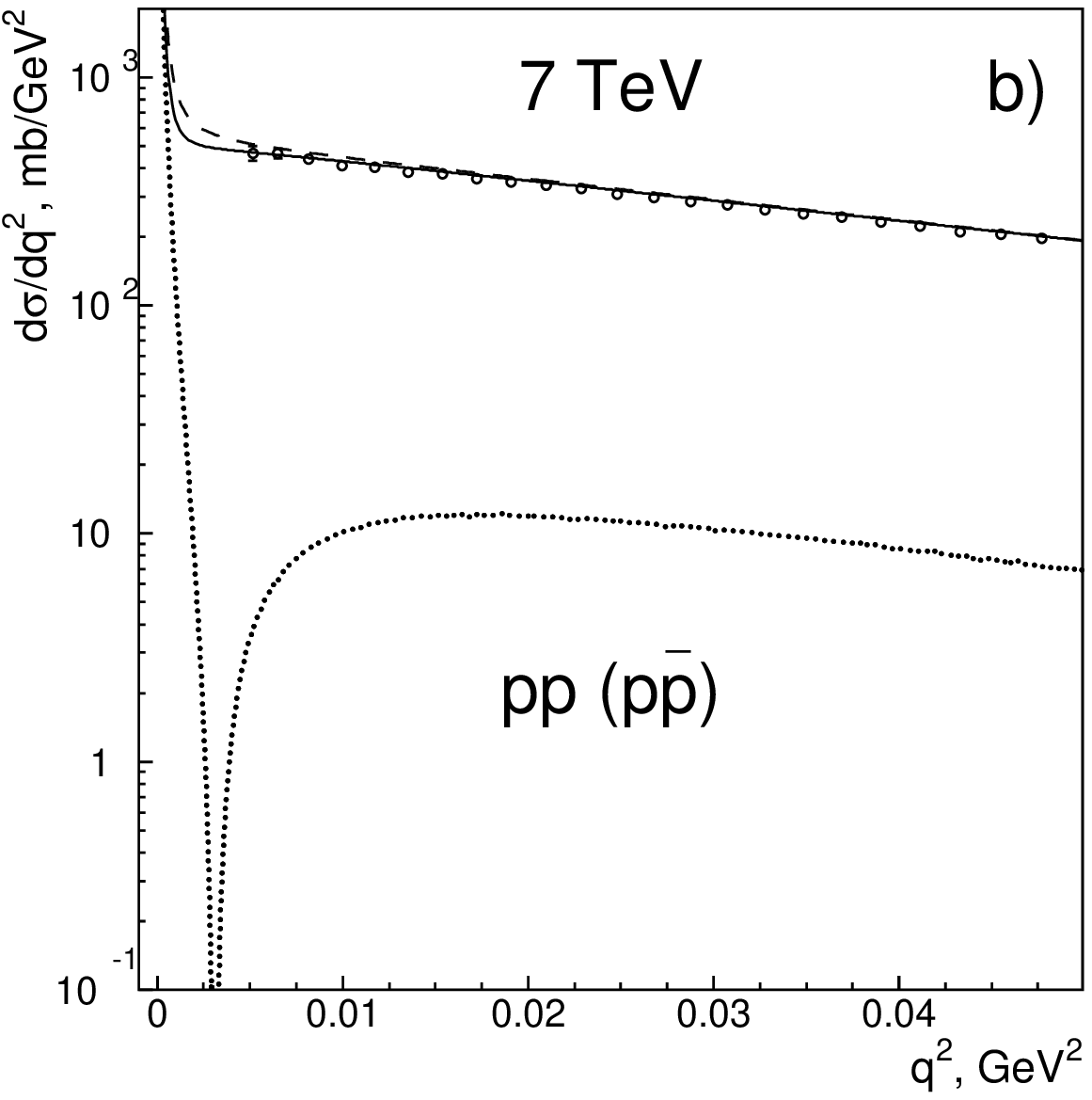,width=0.45\textwidth}}
\centerline{\epsfig{file=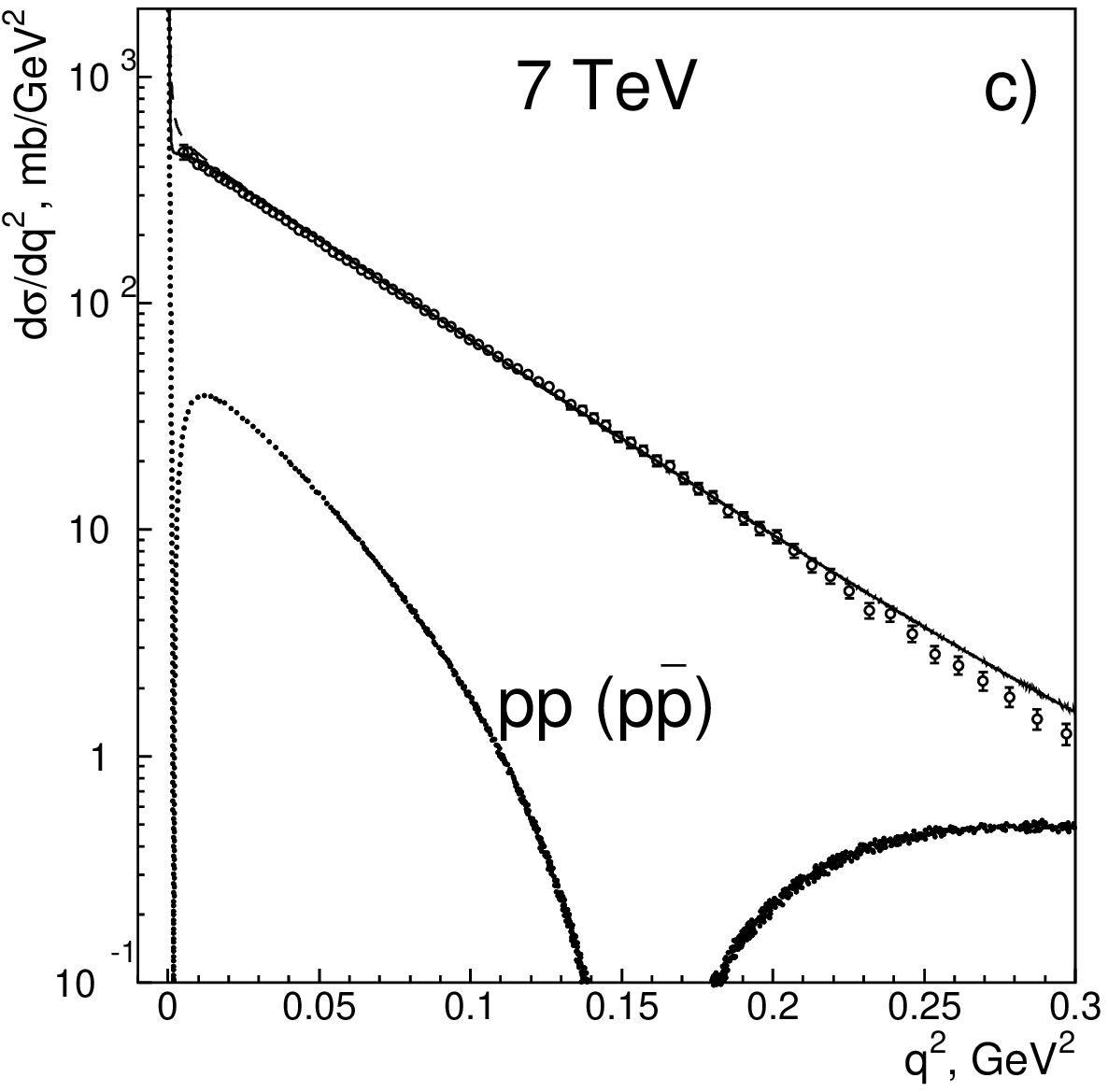,width=0.45\textwidth}
            \epsfig{file=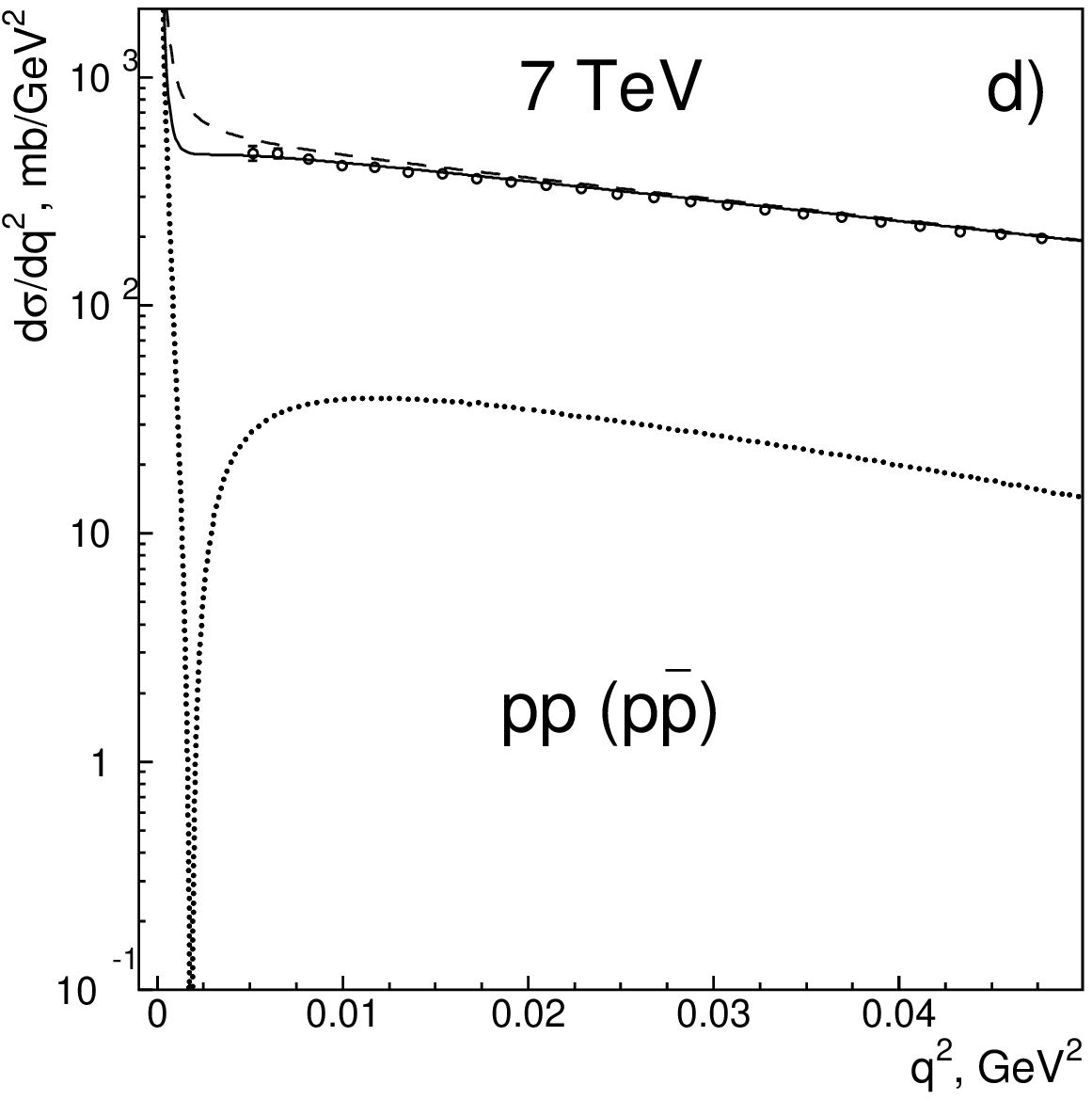,width=0.45\textwidth}}
\caption{
Diffractive scattering cross section for $pp$ at 7 TeV (TOTEM
\cite{Latino:2013ued})
versus description with interplay of hadronic and Coulomb interactions
(Eq. (\ref{e10}), $\lambda$=0.001 GeV): figures (a,b) refer to the black disk mode,
figures (c,d) refer to the resonant disk case;
solid curves refer to $pp$, dashed ones to $p\bar p$.
Dotted curves show a contribution of the real part
in the $pp$ scattering, the last term in Eq. (\ref{e10}).
}
\label{fig2}
\end{figure*}

\begin{figure*}[ht]
%\Fig. 3
\centerline{\epsfig{file=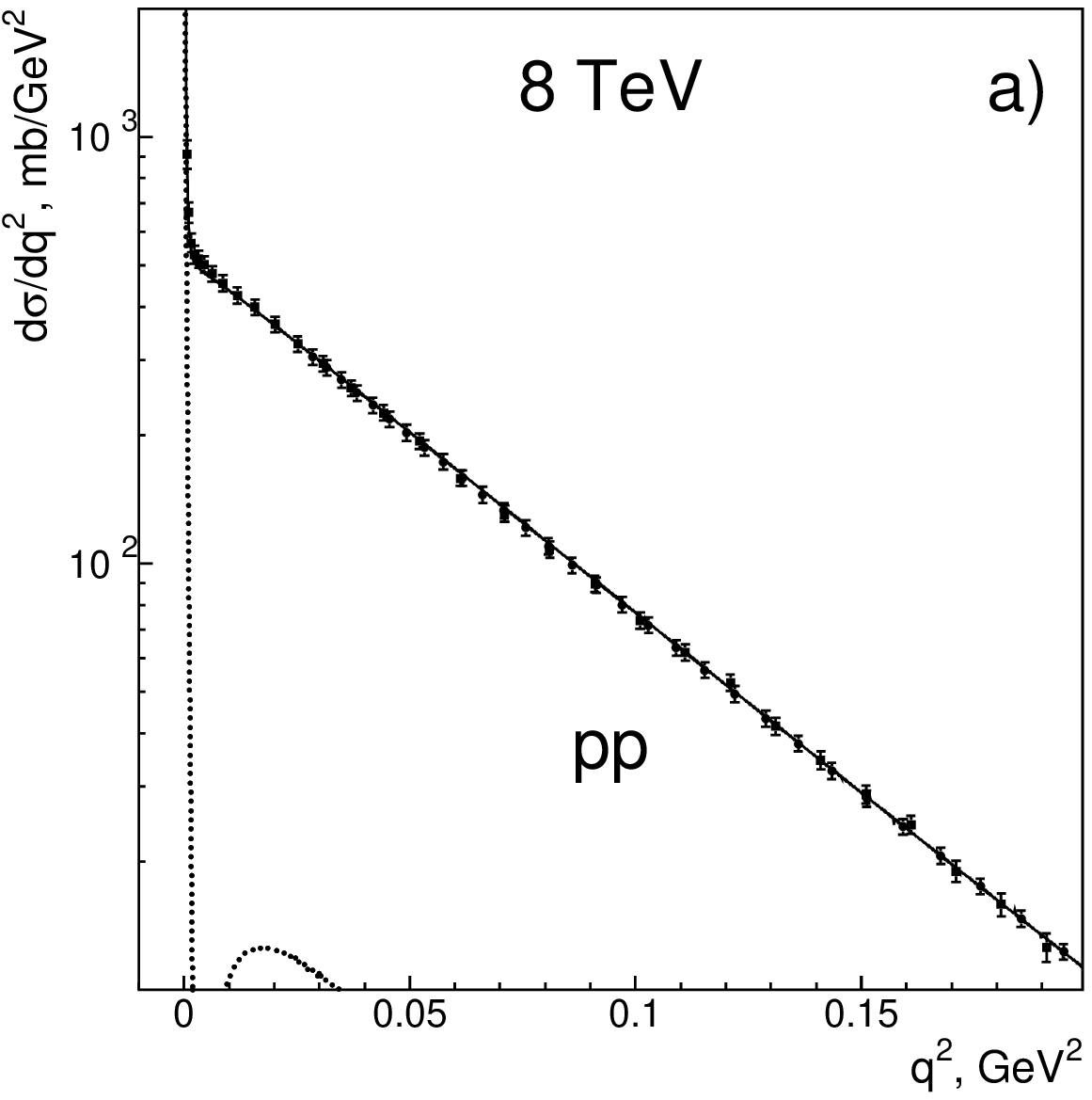,width=0.45\textwidth}
            \epsfig{file=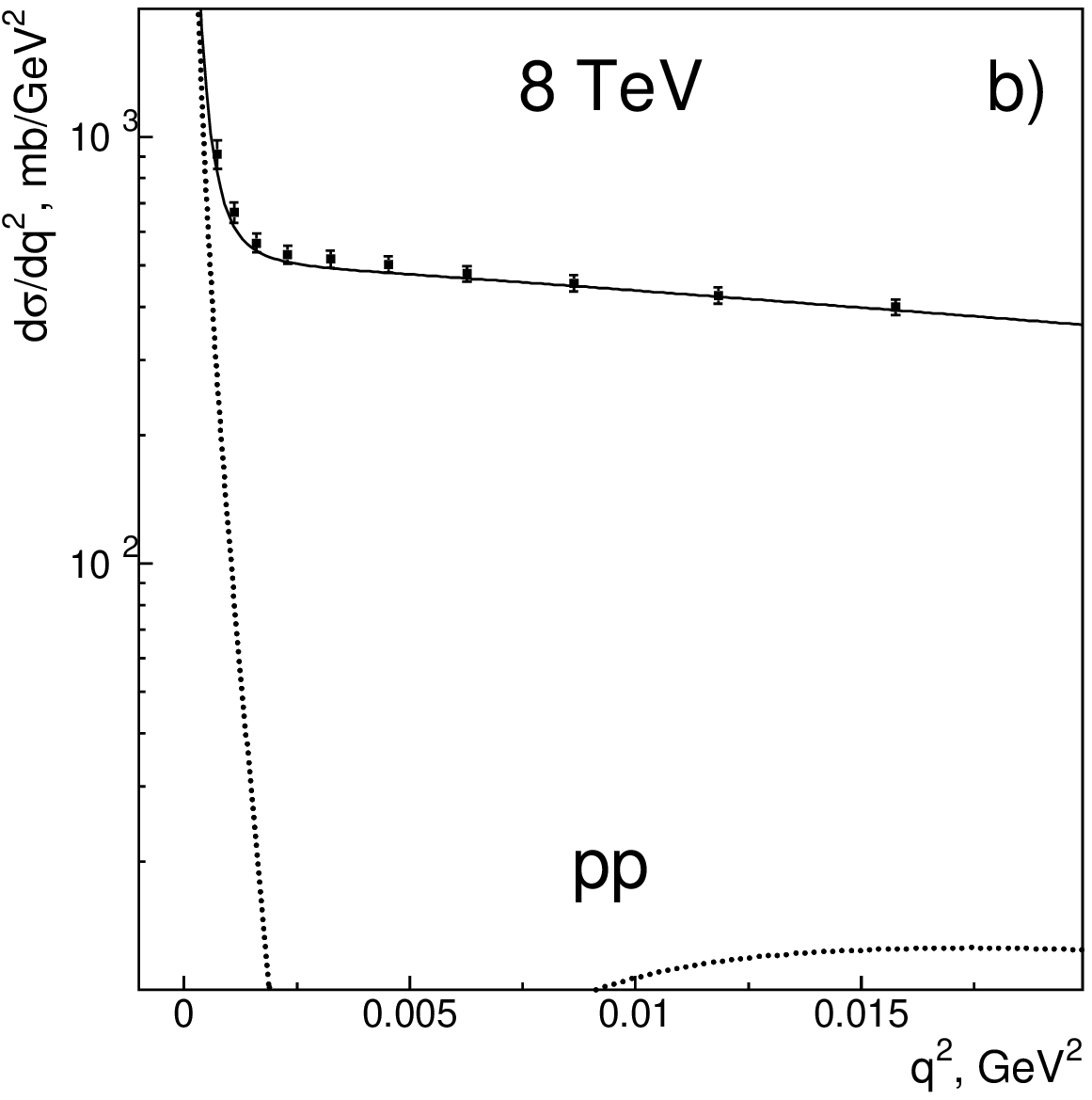,width=0.45\textwidth}}
\centerline{\epsfig{file=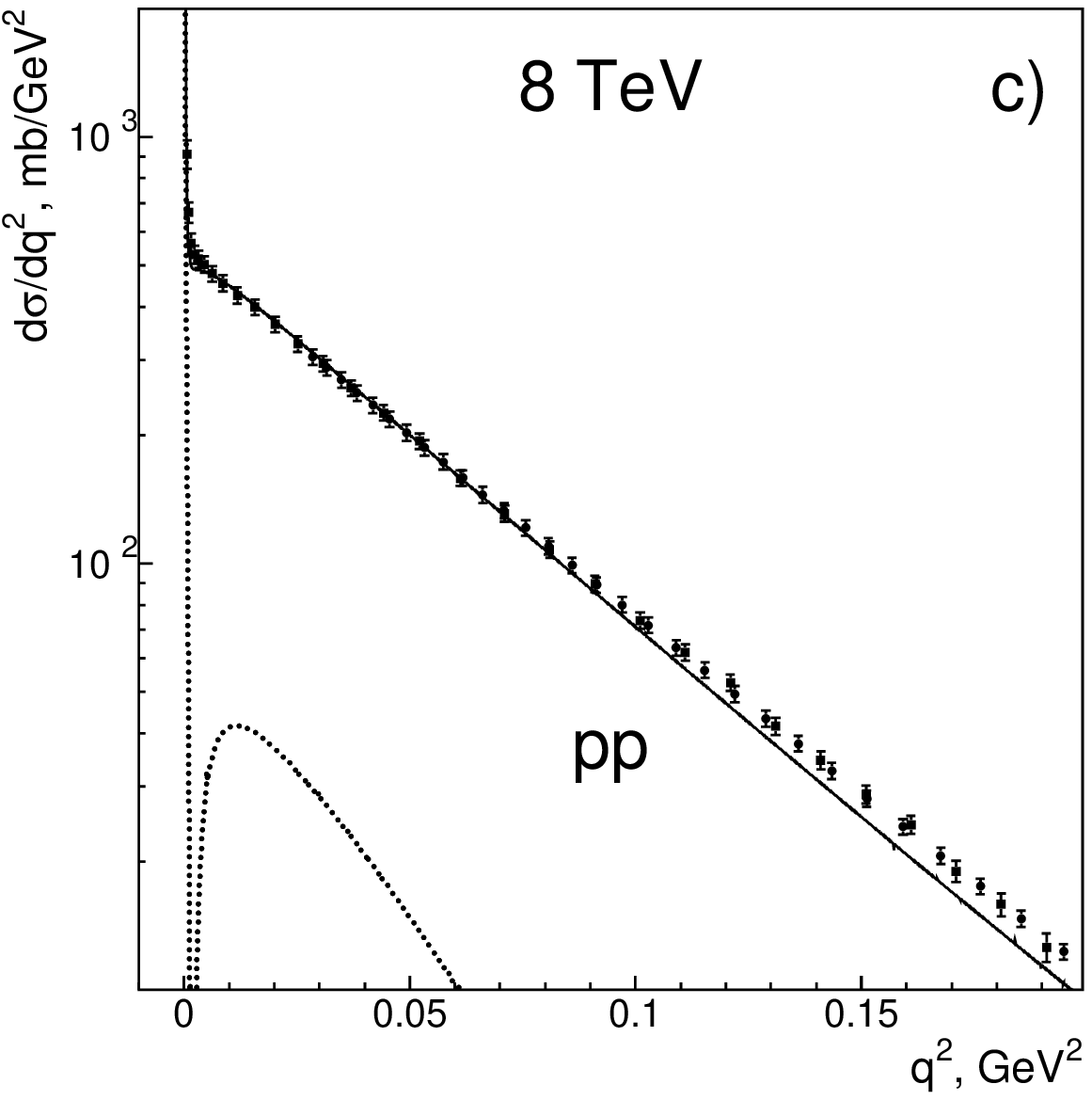,width=0.45\textwidth}
            \epsfig{file=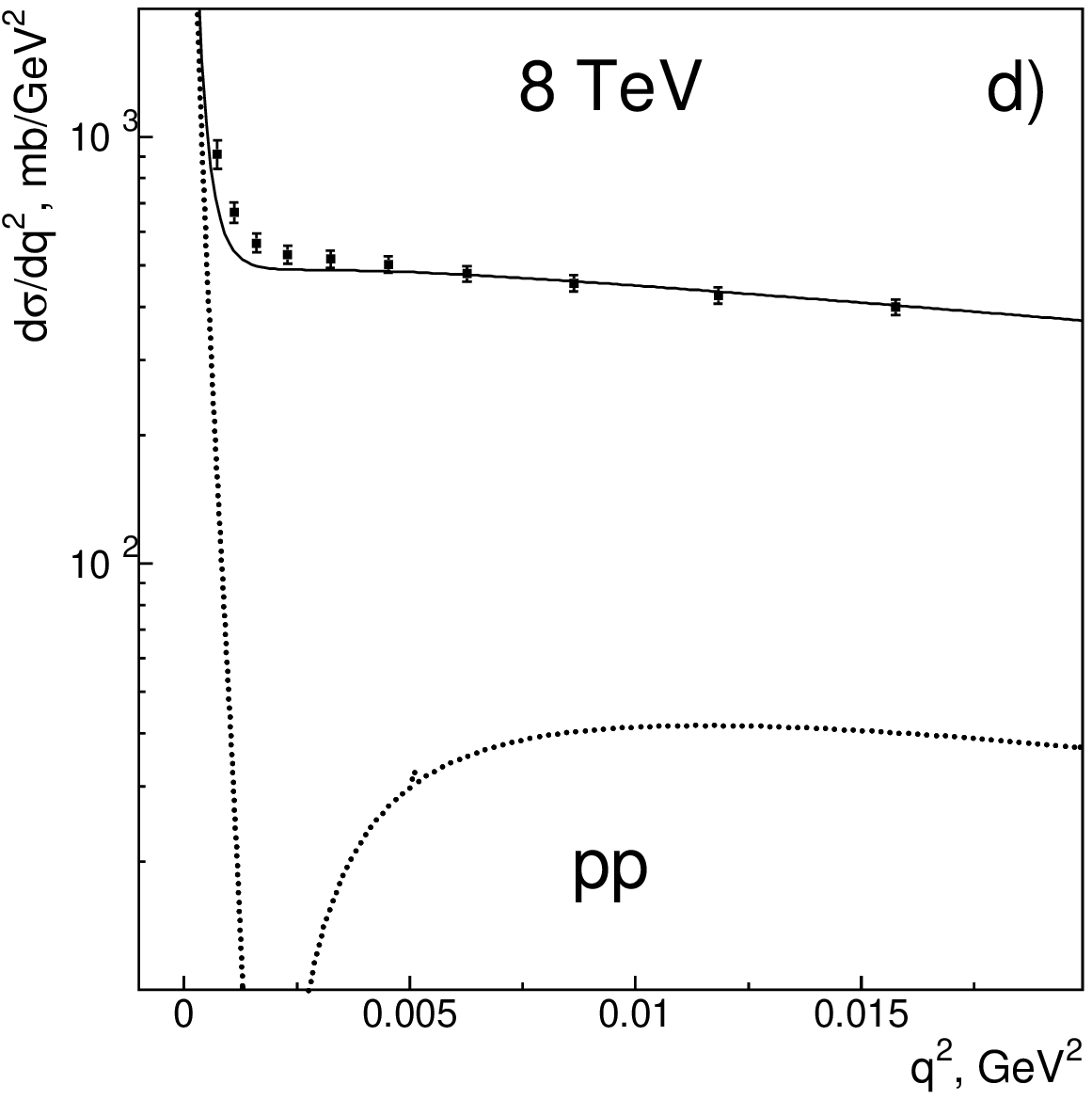,width=0.45\textwidth}}
\caption{
Diffractive scattering cross section for $pp$ at 8 TeV (TOTEM
\cite{8TeV})
versus description with interplay of hadronic and Coulomb interactions
(Eq. (\ref{e10}), $\lambda$=0.001 GeV): figures (a,b) refer to the black disk mode,
figures (c,d) refer to the resonant disk case;
solid curves refer to $pp$.
Dotted curves show a contribution of the real part
in the $pp$ scattering, the last term in Eq. (\ref{e10}).
}
\label{fig3}
\end{figure*}

\begin{figure*}[ht]
%\Fig. 4
\centerline{\epsfig{file=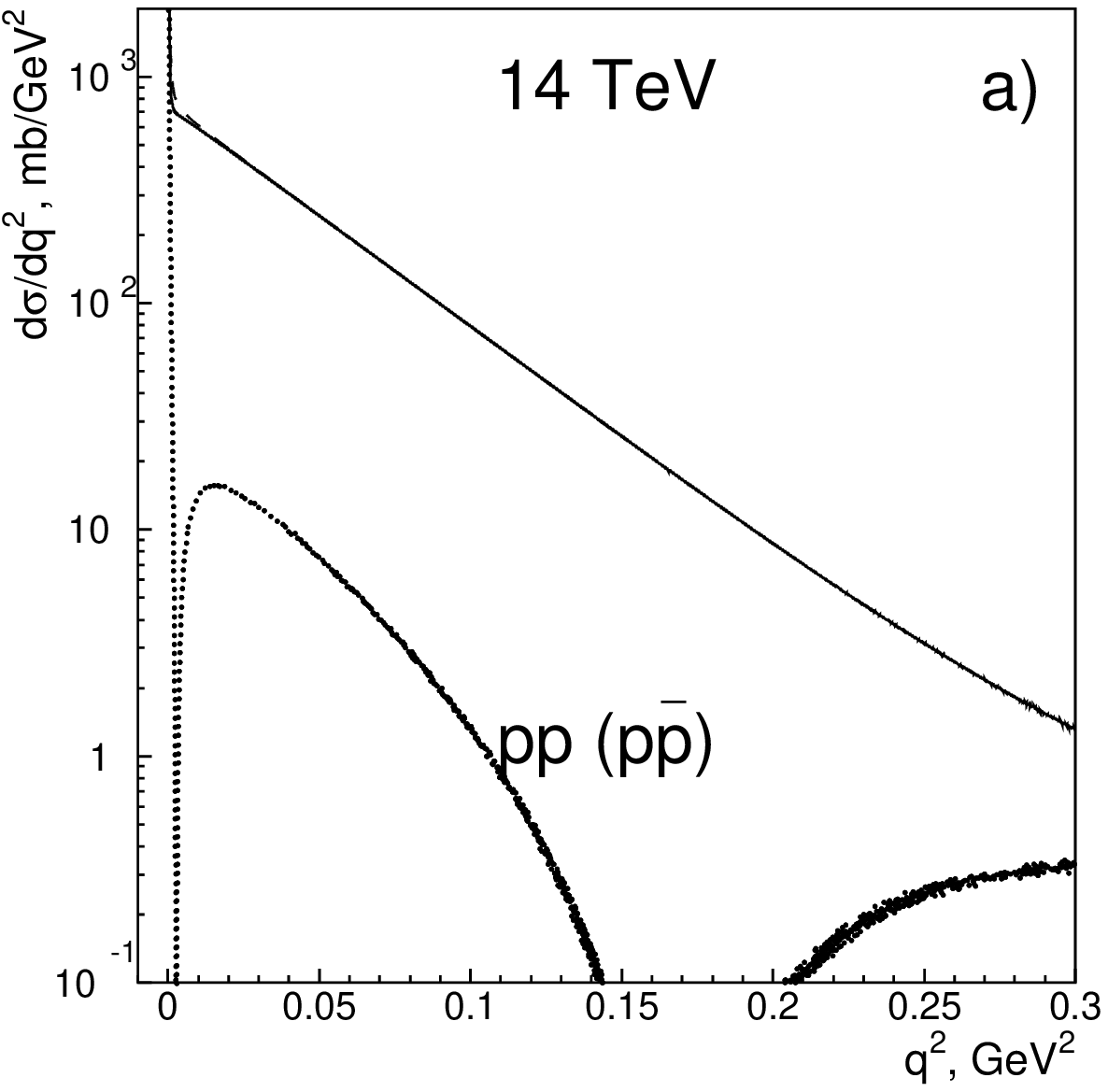,width=0.45\textwidth}
            \epsfig{file=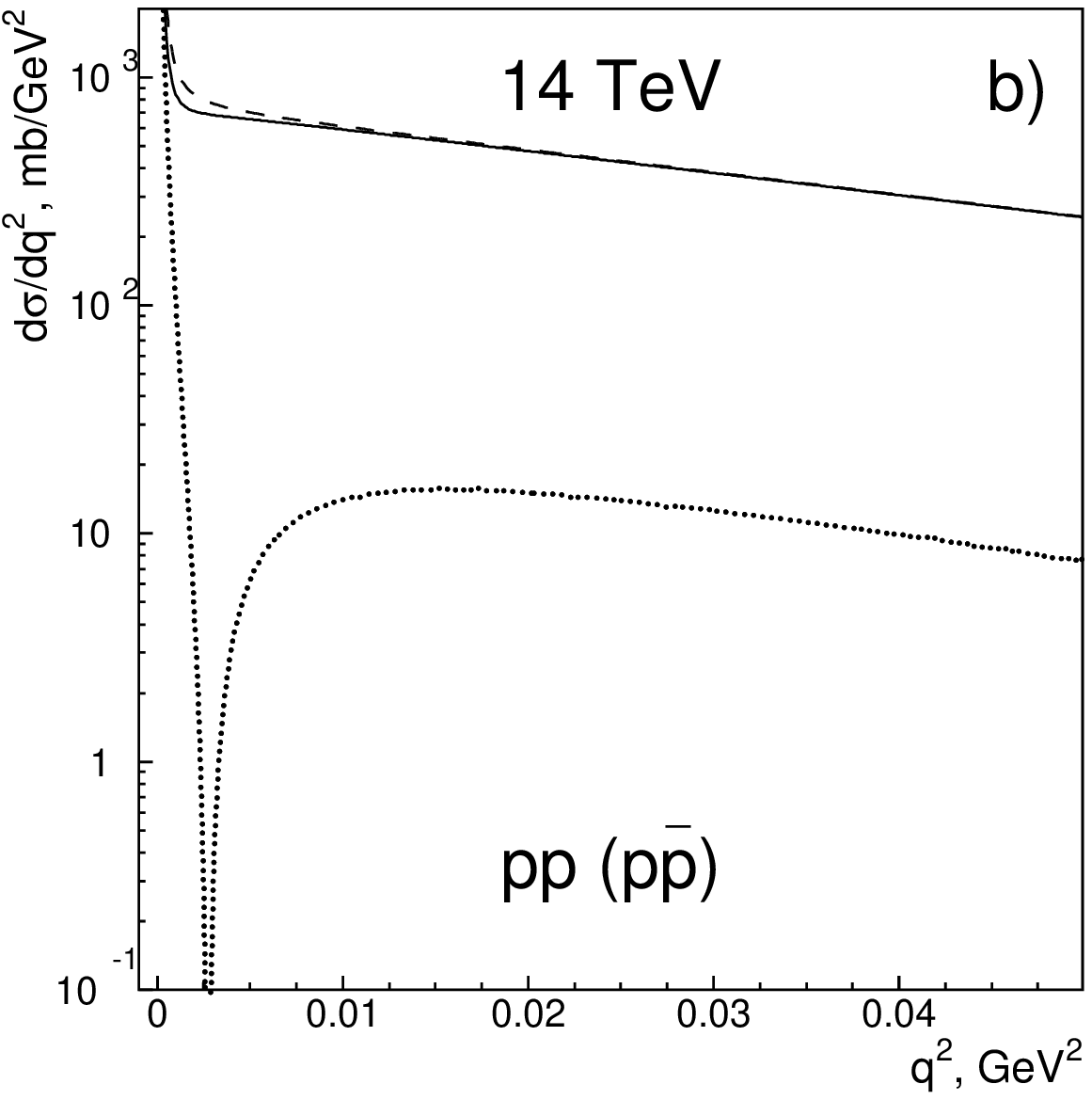,width=0.45\textwidth}}
\centerline{\epsfig{file=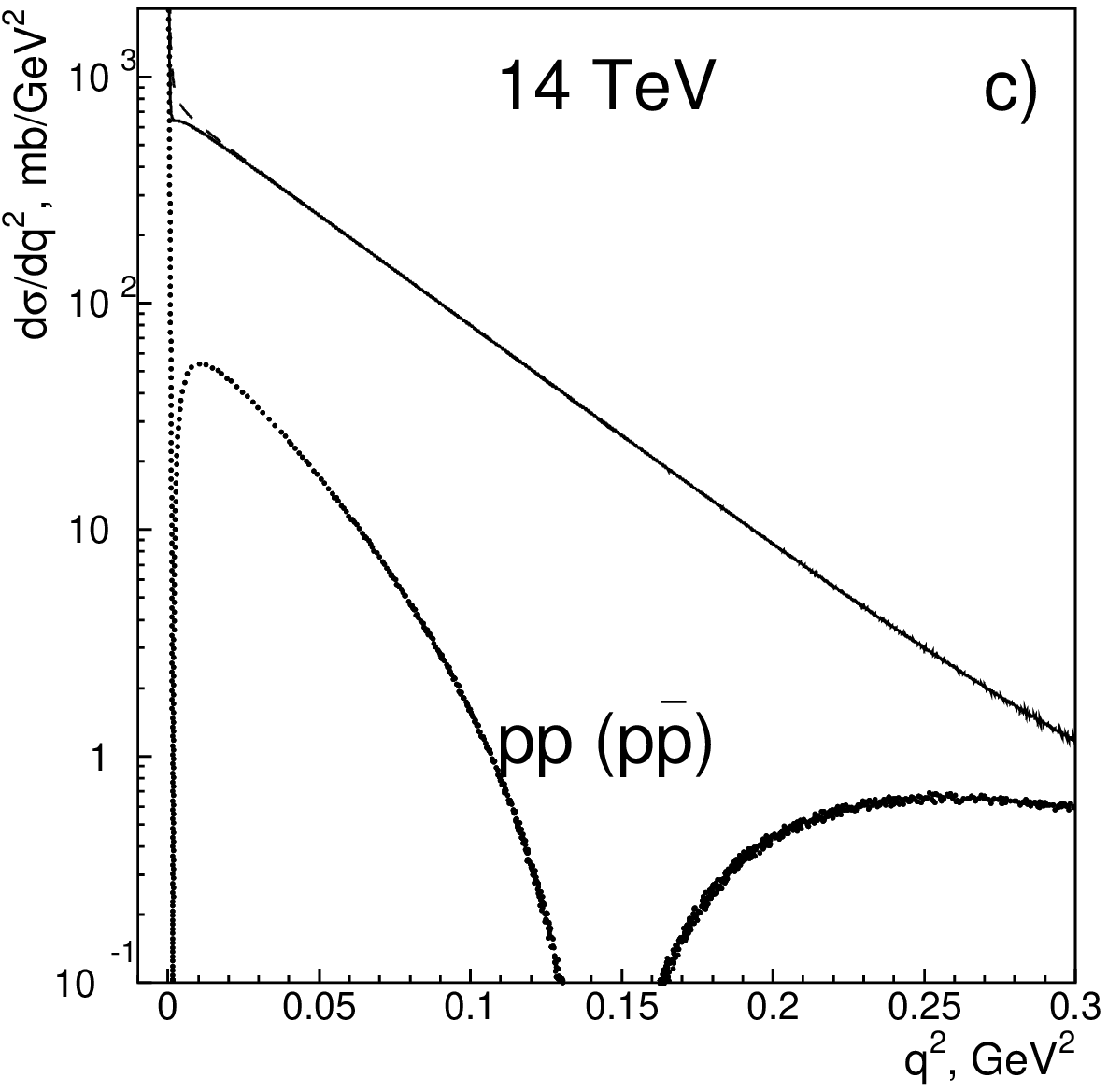,width=0.45\textwidth}
            \epsfig{file=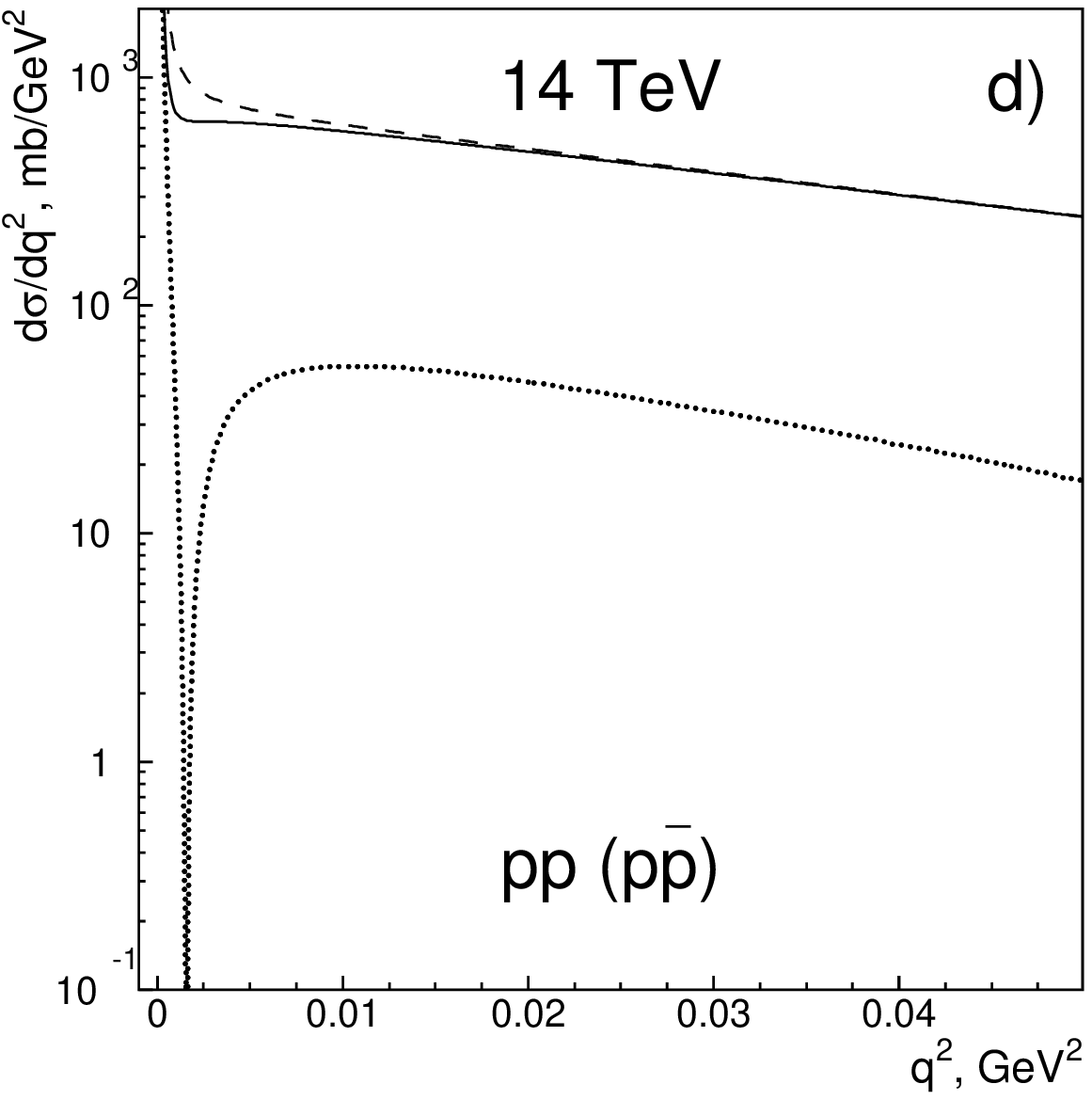,width=0.45\textwidth}}
\caption{
Diffractive scattering cross sections for $pp^\pm$ at 14 TeV
    (Eq. (\ref{e10}), $\lambda$=0.001 GeV):
The figures (a,b) refer to the black disk mode and (c,d) to the resonant disk one;
solid curves refer to $pp$, dashed ones to $p\bar p$.
Dotted curves show a contribution of the real part
in the $pp$ scattering, the last term in Eq. (\ref{e10}).
}
\label{fig4}
\end{figure*}

\section {Diffractive scattering amplitude at ultrahigh energy and
Coulomb interaction }

The interplay of hadronic and Coulomb interactions was studied in a set of
papers, see
\cite{Bethe:1958,Soloviev:1966,West:1968,Franco:1973,Cahn:1982,Kundrat:1993,Kaspar:2011}
and references therein. In the region of ultrahigh energies and small
${\bf q}^2$, the $K$-matrix function technique allows to take into
account directly the combined action of hadronic and Coulomb
interactions ($H+C$).

\subsection{Interplay of hadronic and Coulomb interactions in the $K$-matrix
function technique}

We consider two types of scattering amplitudes
and corresponding profile functions:
the amplitude with combined interaction taken into account,
$A^{C+H}({\bf q}^2 ,\xi)$ and $T^{C+H}(b,\xi)$,
and that with the switched-off Coulomb interaction,
$A^H({\bf q}^2 ,\xi)$ and $T^H(b,\xi)$.

For the combined
interaction profile function we write:
\bea \label{6}
    T^{C+H}(b,\xi)=
\frac{-2iK^{C+H}(b,\xi)}{1-iK^{C+H}(b,\xi)}
=
\frac{-2i\left(K^{C}(b)+K^H(b,\xi)\right)}{1-i\left(K^{C}(b)
+K^H(b,\xi)\right)}\,,
\eea
where the Coulomb interaction is written as:
\bea
\label{13}
A^C({\bf q}^2 )&=&
\pm i f_1({\bf q}^2)\frac{4\pi\alpha}{{\bf q}^2+\lambda^2}
f_2({\bf q}^2),
\nn \\
-2i K^C(b)&=&
\pm i\int\frac{d^2q}{(2\pi)^2 }  e^{i{\bf q}{\bf b}}
f_1({\bf q}^2)
\frac{4\pi\alpha}{{\bf q}^2+\lambda^2} f_2({\bf q}^2)\,.
\eea
Here  $\alpha =1/137$; the upper/lower signs refer to the same/opposite charges
of the colliding particles.
The cutting parameter $\lambda$, which removes the infrared divergency,
can tend to zero
in the final result for $A^{C+H}({\bf q}^2 ,\xi)$.
Colliding hadron form factors, $f_1({\bf q}^2)$ and $f_2({\bf q}^2)$,
guarantee the
convergence of the integrals at ${\bf q}^2\to \infty$;
for the $pp^\pm$ collisions we use:
\be
f_1({\bf q}^2)=f_2({\bf q}^2)=\frac{1}{(1+\frac{{\bf q}^2}{0.71GeV^2})^2} \,.
\ee
In Fig. {\ref{fig1}} we show $-K^{C}(b) $
for $\lambda =0.01$ GeV (Fig. {\ref{fig1}}a) and $0.001$ GeV
(Fig. {\ref{fig1}}b).  At large
$b$ the $K^C(b)$ is a scaling function in terms of $\beta=\lambda b$,
the $\beta$-scaling gets broken at small $\beta$.

For a diffractive scattering cross section with Coulomb interaction taken into
account we write now:
   \bea \label{e10}
A^{H+C}({\bf q}^2 ,\xi)&=& \int d^2b e^{i{\bf q}{\bf
b}}T^{H+C}(b,\xi),   \nn
\\
4\pi\frac{d\sigma_{el}}{d{\bf q}^2}&=& \Big|A_{\Im}^{H+C}({\bf
q}^2)\Big|^2+\Big|A_{\Re}^{H+C}({\bf q}^2)\Big|^2\,.
\eea
The diffractive scattering cross sections with Coulomb interaction taken
into account are shown in Figs. \ref{fig2}, \ref{fig3}, \ref{fig4} for $\sqrt s=$7, 8, 14
TeV.

\subsection{Interference of hadronic and Coulomb interactions  }

\begin{figure*}[ht]
%Fig.5
\centerline{\epsfig{file=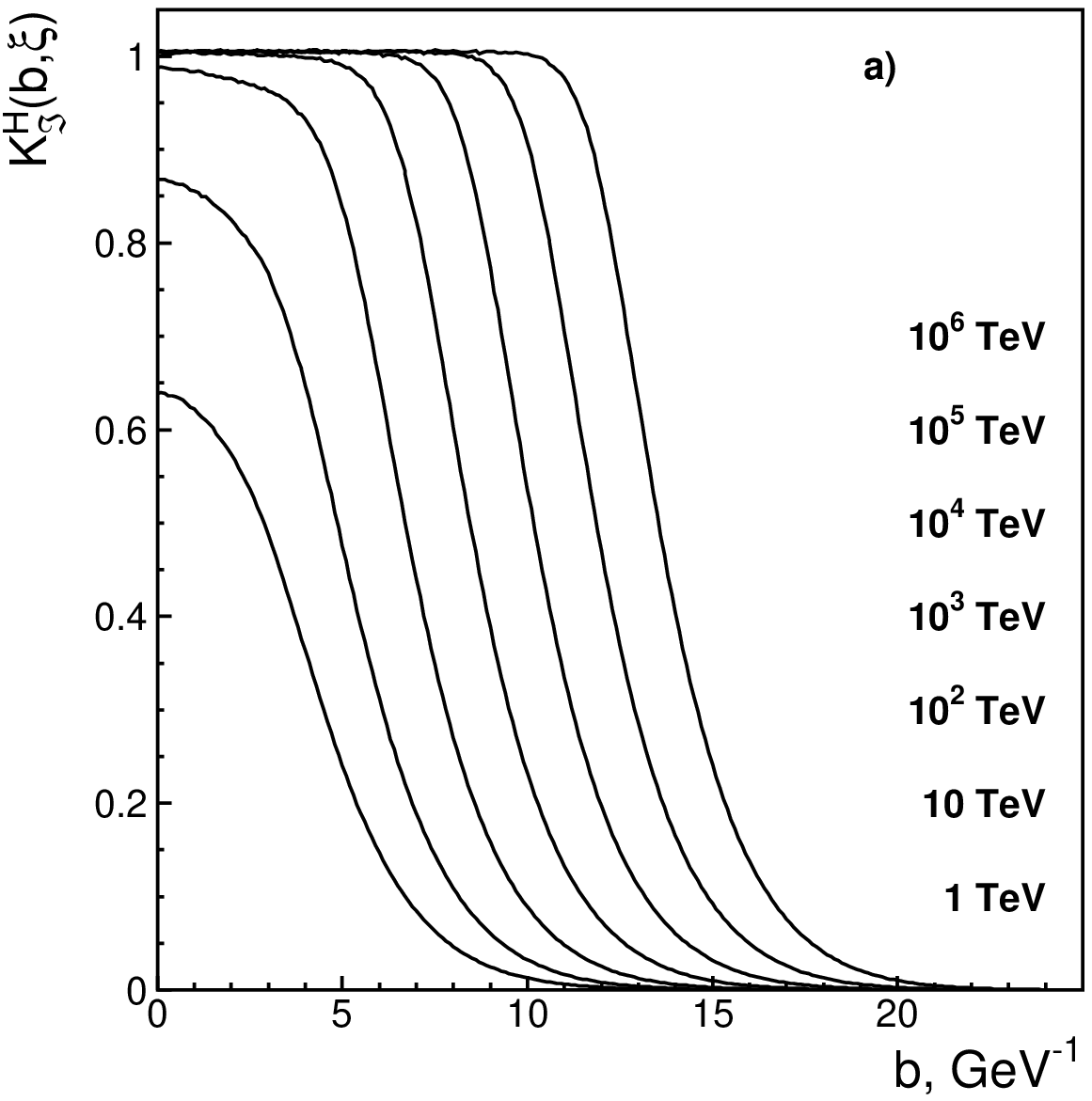,width=0.45\textwidth}
            \epsfig{file=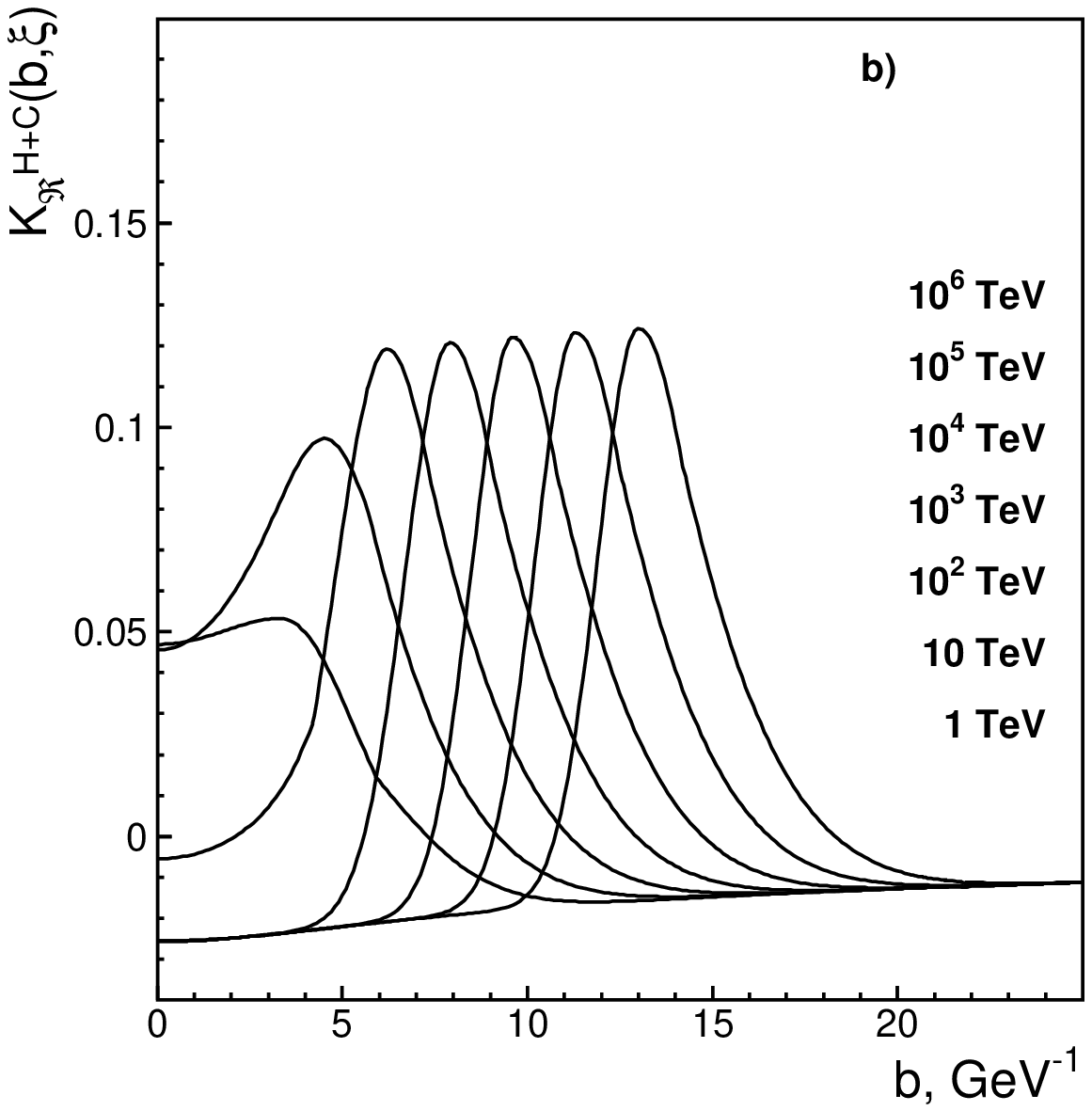,width=0.45\textwidth}}
\centerline{\epsfig{file=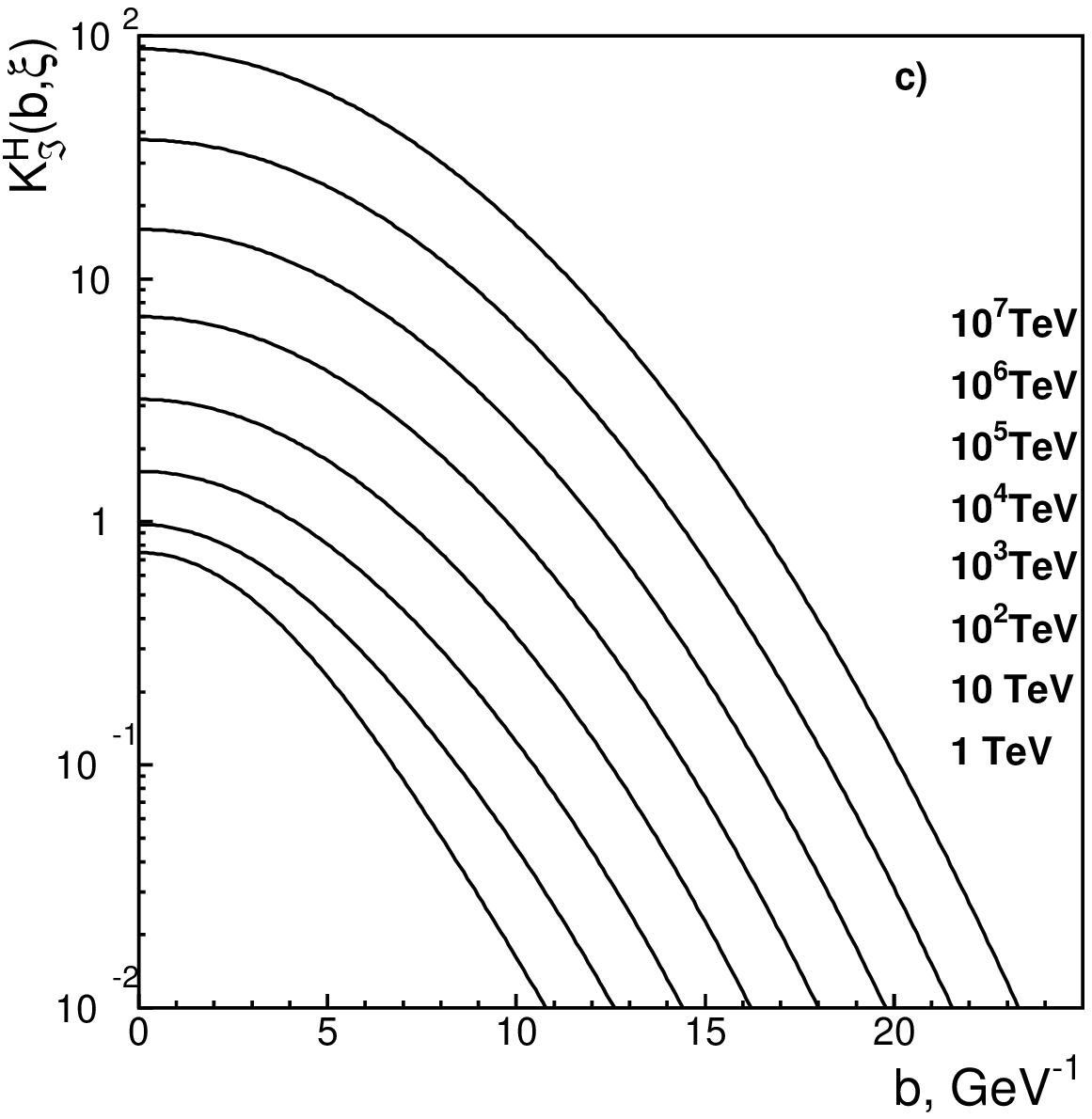,width=0.45\textwidth}
            \epsfig{file=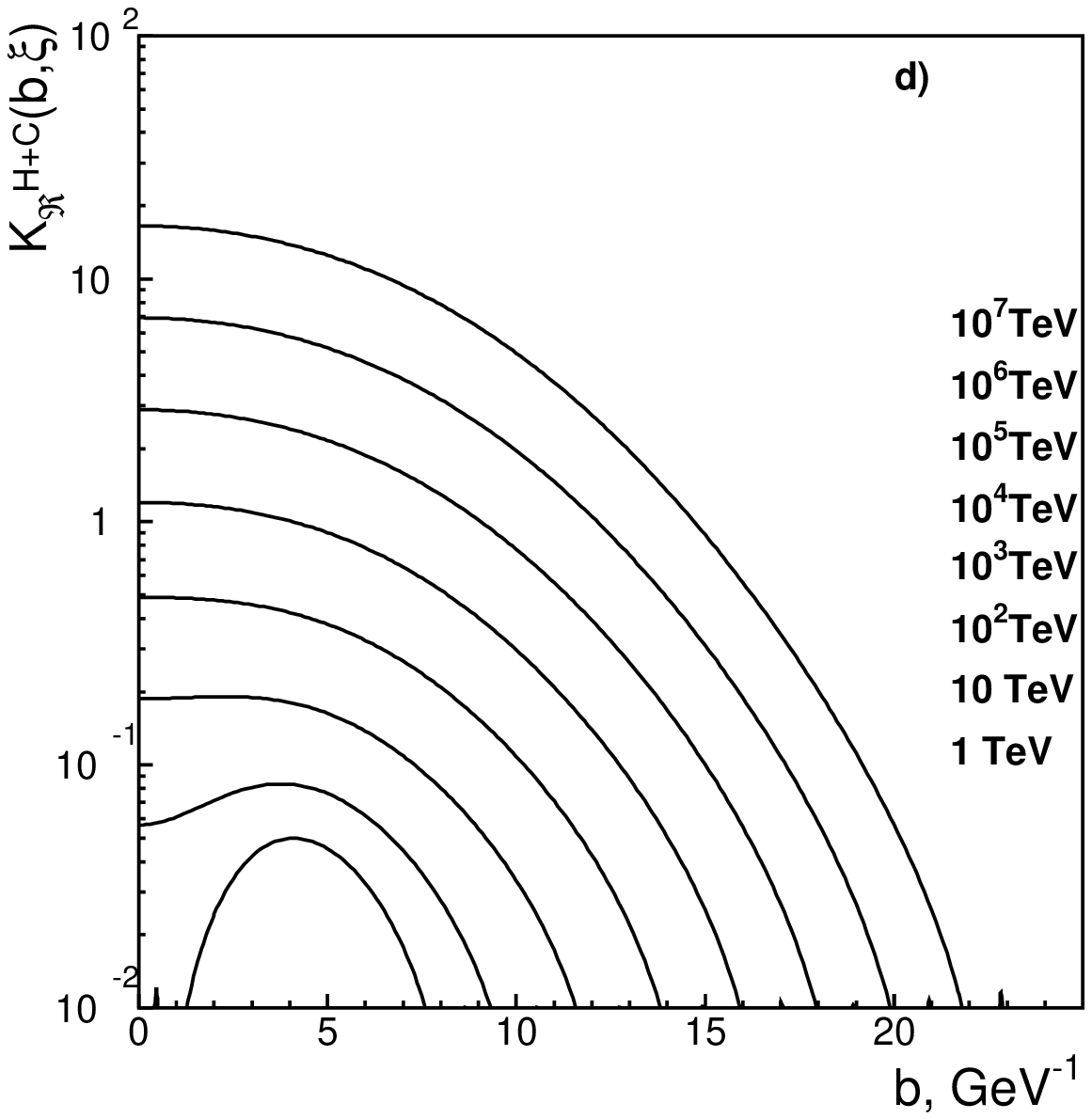,width=0.45\textwidth}}
\caption{
K-matrix functions for $pp$ scattering
 (imaginary and real parts) in the energy region $\sqrt s=1-10^6$ TeV;
figures (a,b) refer to the black disk mode,
(c,d) to the resonant disk mode.
}
\label{fig5}
%\label{fig_kmatr}
\end{figure*}

\begin{figure*}[ht]
%Fig.6
\centerline{\epsfig{file=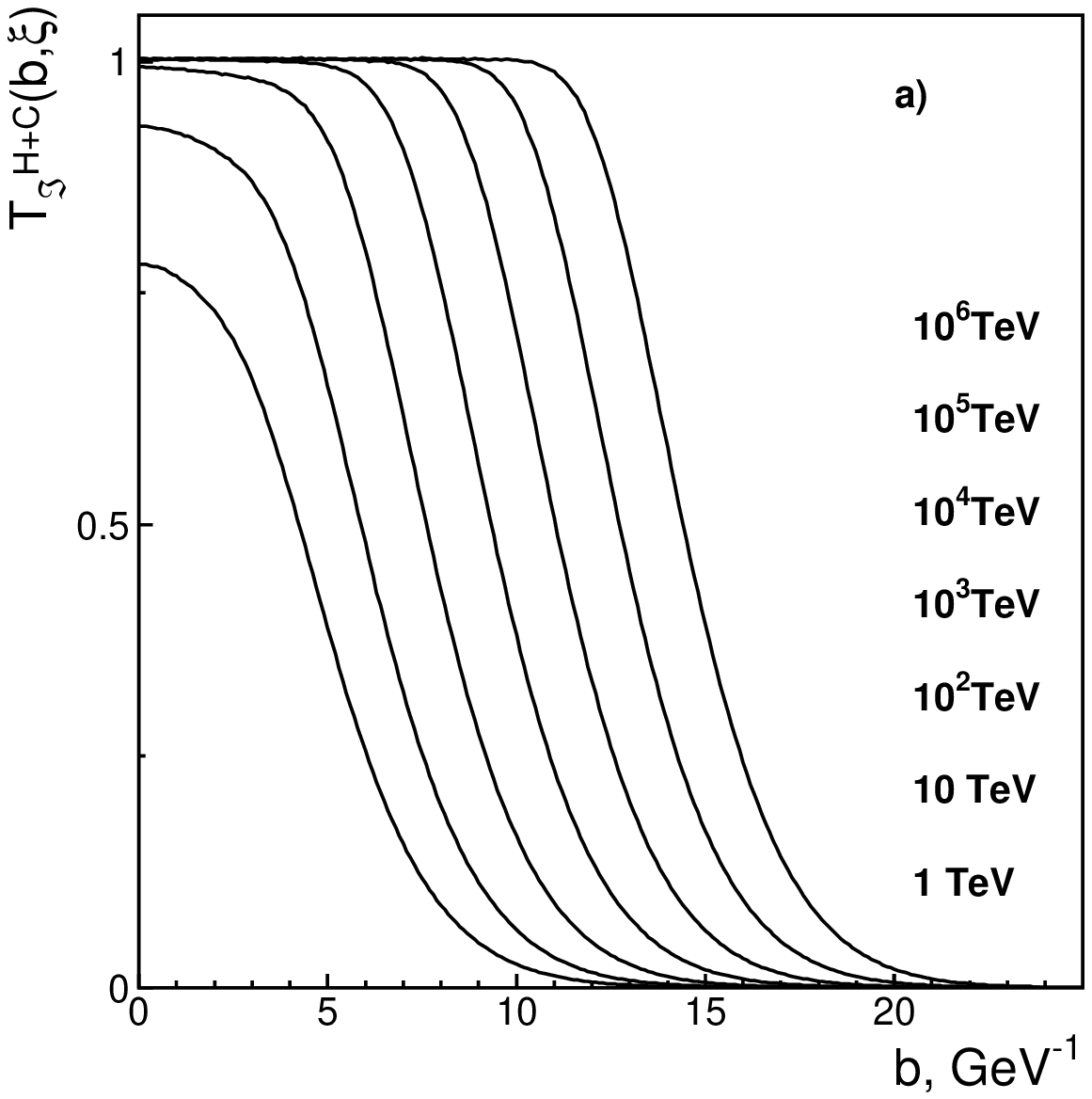,width=0.45\textwidth}
            \epsfig{file=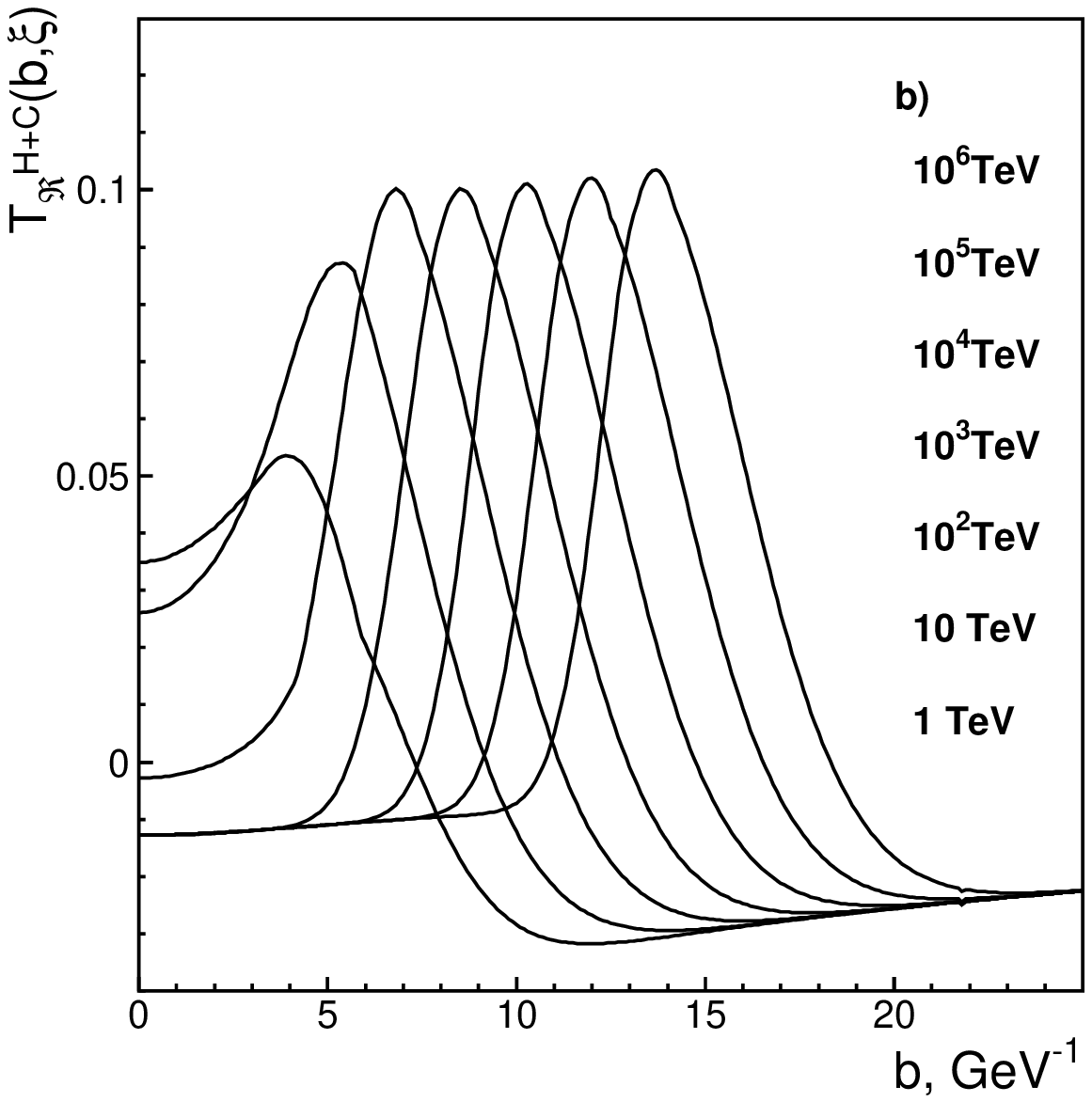,width=0.45\textwidth}}
\centerline{\epsfig{file=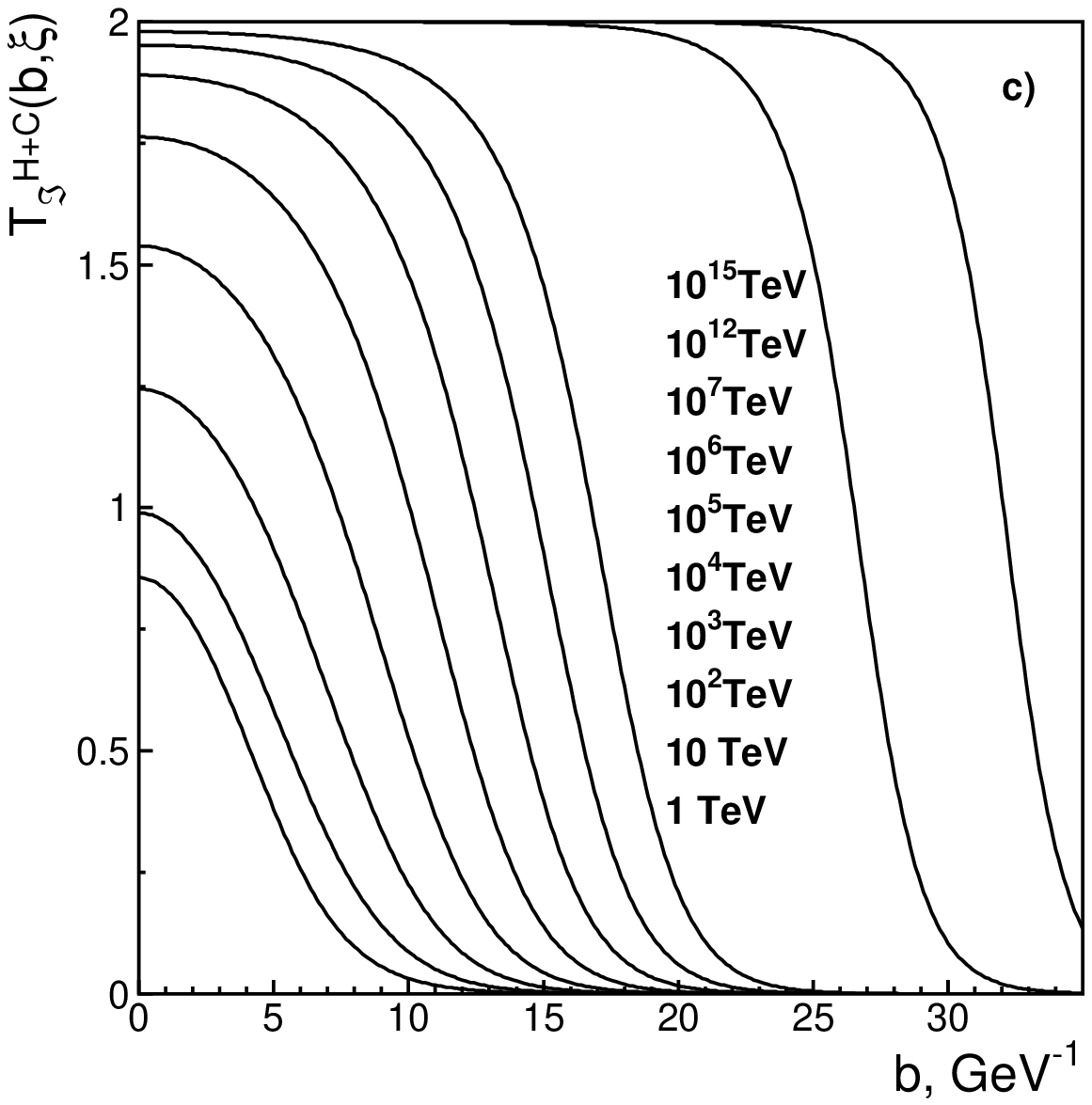,width=0.45\textwidth}
            \epsfig{file=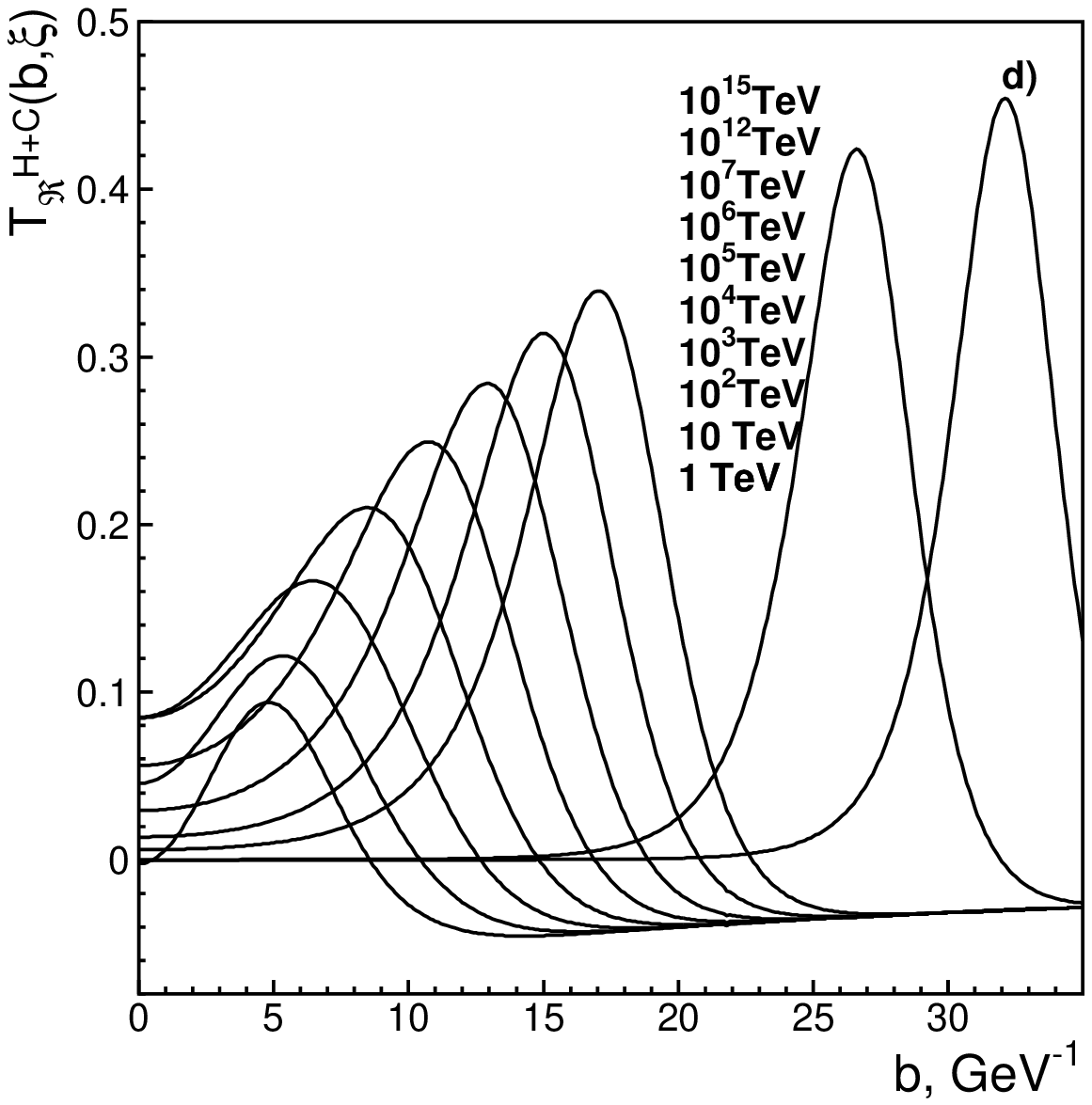,width=0.45\textwidth}}
\caption{
Profile functions for $pp$ scattering
(imaginary and real parts) in the energy region $\sqrt s=1-10^6$ TeV
with the Coulomb interaction taken into account;
(a,b) refer to the black disk mode,
(c,d) to the resonant disk mode.}
\label{fig6}
%\label{fig_profile}
\end{figure*}

At the LHC energies ($\sqrt s \sim 10$ TeV) the real part of the
resonant disk amplitude demonstrates an essentially stronger contribution
compared to that in the black disk mode, see Figs. \ref{fig2},
\ref{fig3}, \ref{fig4} for  $\Big|A_{\Re}^{H+C}({\bf q}^2)\Big|^2$ and
Figs. \ref{fig5}, \ref{fig6} for  $K^{H+C}(b,\xi)$ and $T^{H+C}(b,\xi)$. For
the resonant disk mode it results in a shoulder in the $pp$ diffractive
cross section $\frac{d\sigma_{el}({\bf q}^2)}{d{\bf q}^2}$ at
${\bf q}^2\sim 0.0025-0.0075$ GeV$^2$.
The shoulder is a qualitative attribute of the resonant disk picture in
the $pp$ scattering, and it means that a presence of the shoulder, or
its absence, can serve as a sign for the disk modes.

\subsection{Asymptotic regimes in the black disk and resonant disk modes}

For the black disk mode the valuable asymptotic regime works
at $\sqrt s \ga 10^2$ TeV. The real parts of $K^{H+C}(b,\xi)$ and
$T^{H+C}(b,\xi)$ (see  Figs. \ref{fig5}d, \ref{fig6}d) have peak
shapes, with fixed maximal peak values $\sim$0.12 and $\sim$0.10,
respectively. At large $b$ where the pure Coulomb interaction works
($b> R_{disk\,radius}$) the $K$-matrix function (and the profile
function) is negative for the $pp$ scattering as it should be for
particles with the same electric charge.

In the resonant disk mode the asymptotic value of the imaginary part of
the amplitue reaches its maximal value, $T_\Im^{H+C}(b,\xi)\simeq 2$
at $\sqrt s \ga 10^5$ TeV only. The maximal
values of the real part continue to increase at  $\sqrt s \ga 10^6$
TeV. The stable maximal values are reached at $\sqrt s \ga 10^{19}$ TeV
where $T_\Re^{H+C}(maximal)\simeq 0.6$. Negative values of the real
part of the $pp$ scattering profile function, $T_\Re^{H+C}(b,\xi)<0$ at
$b> R_{disk\,radius}$, are inherent to resonant disk mode as well.

\begin{figure*}[ht]
%\Fig. 7
\centerline{\epsfig{file=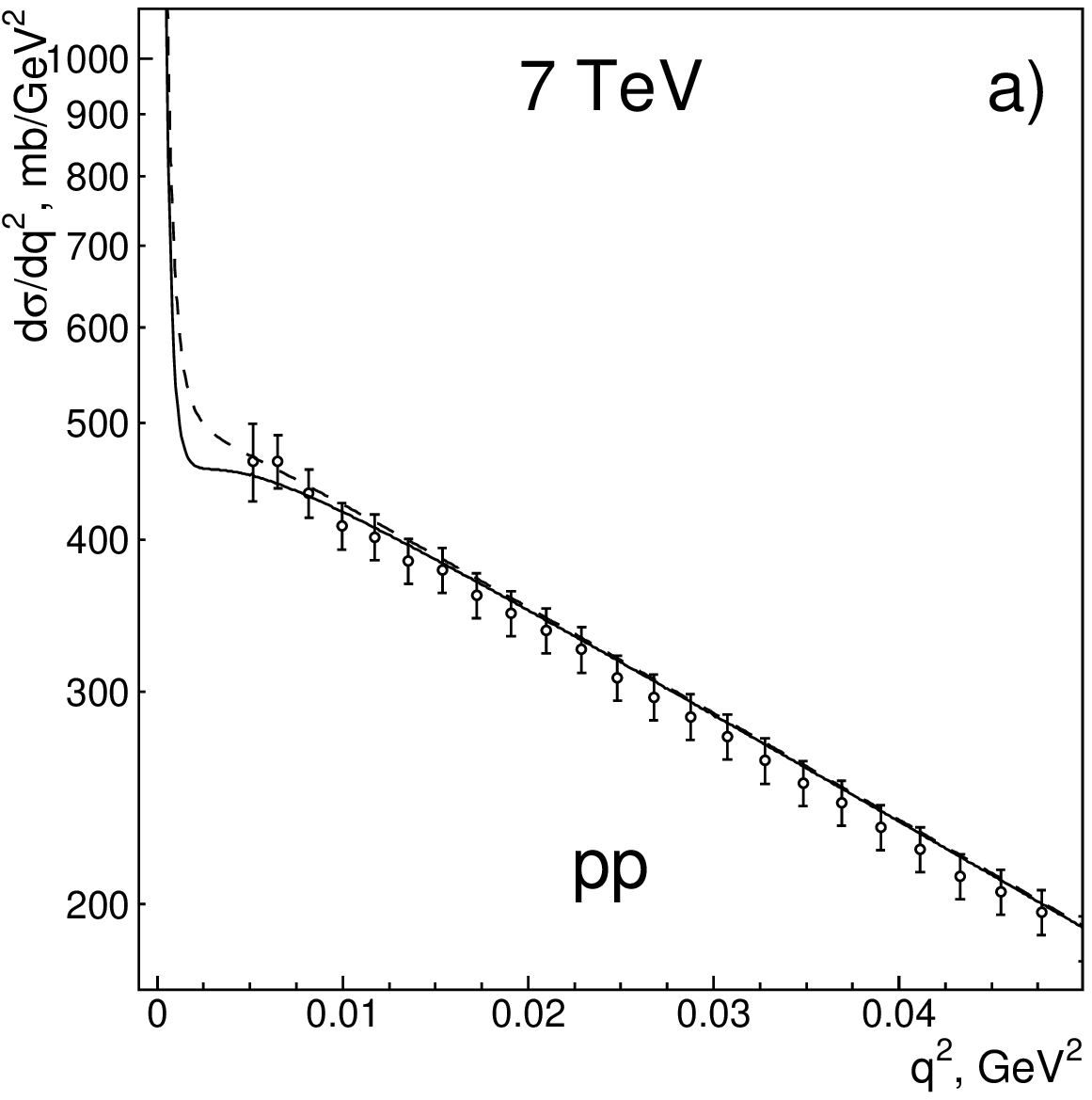,width=0.45\textwidth}
            \epsfig{file=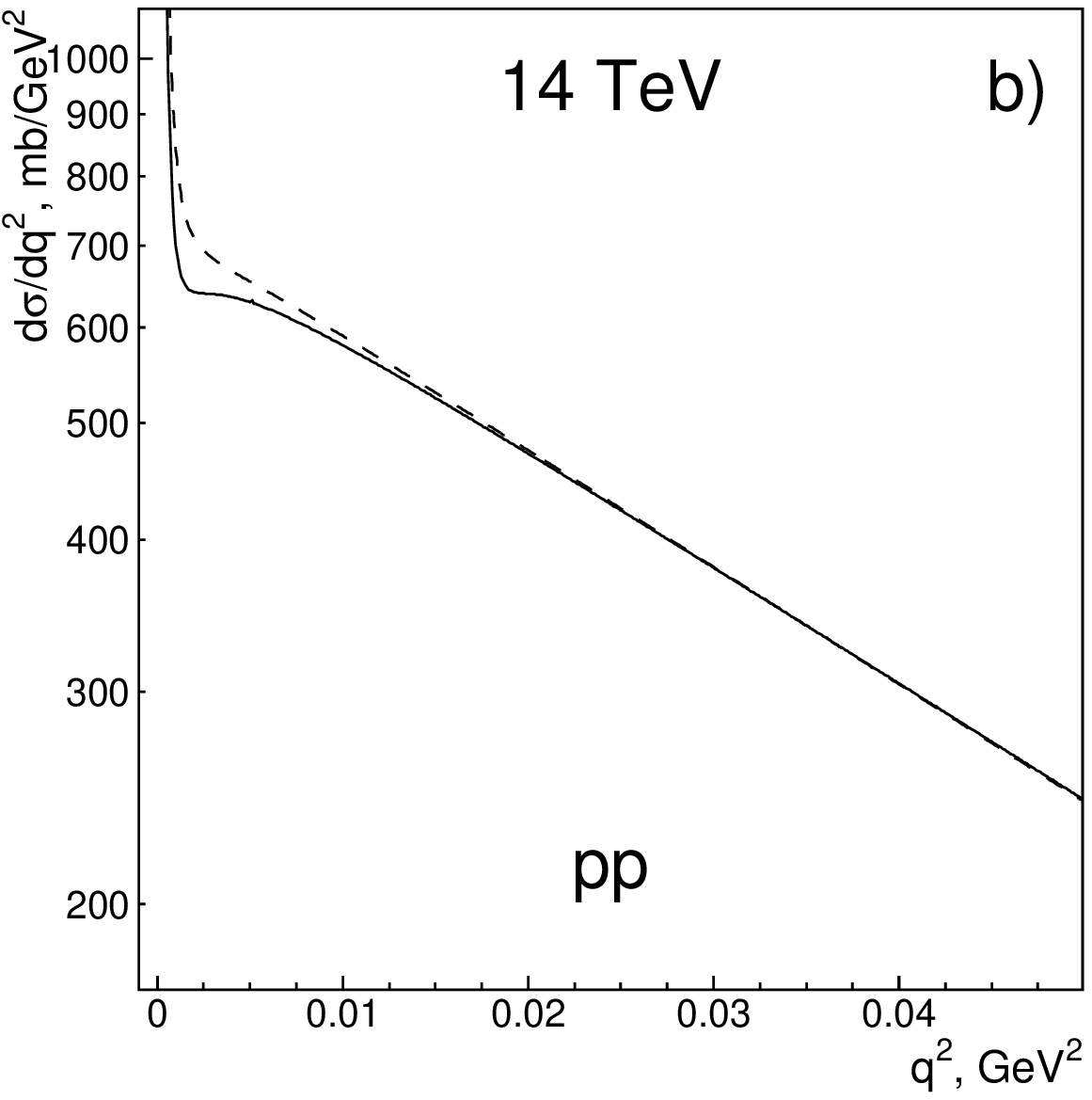,width=0.45\textwidth}}
\caption{
Diffractive scattering cross sections for $pp$ at 7 TeV (a)
 (Eq. (\ref{e10}), $\lambda$=0.001 GeV).
The dashed line refers to the black disk mode and the solid one
to the resonant disk mode.}
   \label{fig7}
   \end{figure*}

\subsection{The 8 TeV data as an argument for the black disk mode}

The resonant disk picture at $\sqrt s \sim 10$ TeV results in an essentially
larger value of the real part in the scattering amplitude
than that in the black disk mode. The large real part is realized in a shoulder
in the $pp$ diffractive cross section at
${\bf q}^2\sim 0.0025-0.0075$ GeV$^2$.  The shoulder is absent in
the data at 8 TeV \cite{8TeV} that agree with calculations for the black disk
picture.

Figures \ref{fig7} demonstrate the region of small momenta transferred
for the $pp$ diffractive cross sections at 7 TeV (the data from
ref. \cite{Latino:2013ued})
and 14 TeV, the solid curves refer to the resonant disk mode and the dashed curve to the
black disk one. The data \cite{Latino:2013ued} do not distinguish the modes being well
described in both of them.

\begin{figure*}[ht]
%\Fig. 8
\centerline{\epsfig{file=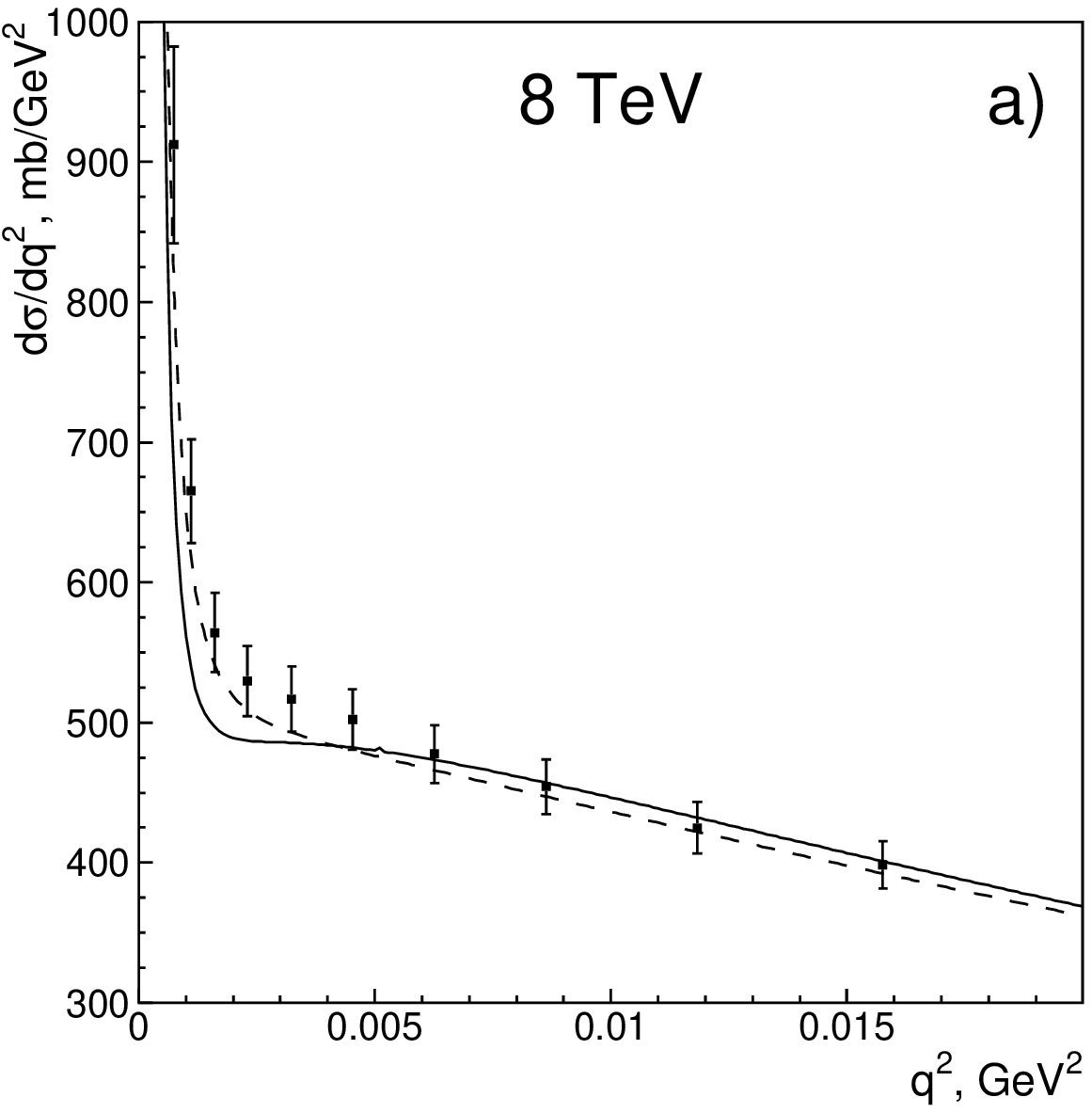,width=0.45\textwidth}
            \epsfig{file=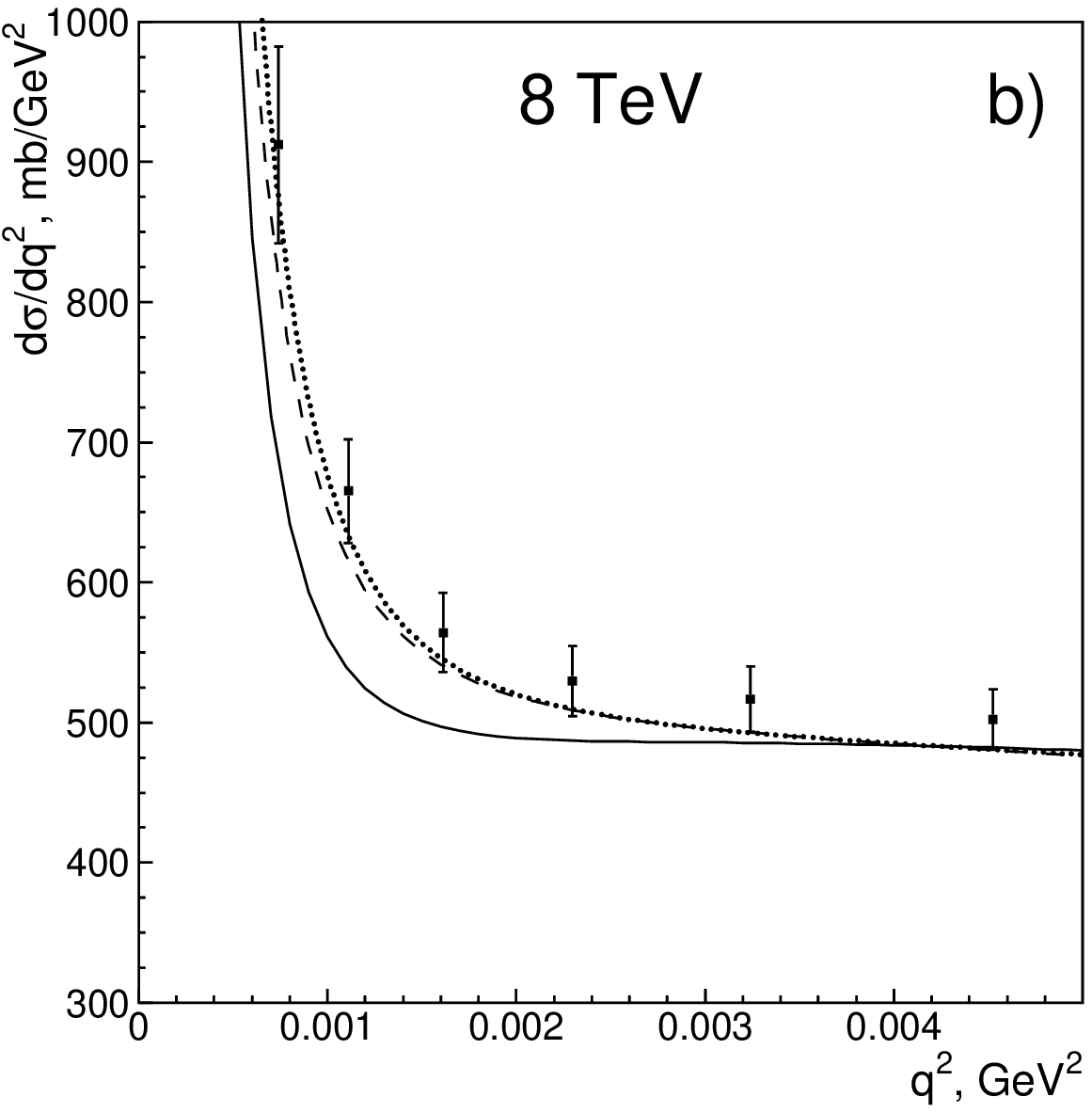,width=0.45\textwidth}}
\caption{
Diffractive scattering cross sections for $pp$ at 8 TeV
   (Eq. (\ref{e10}), $\lambda$=0.001 GeV).
The dashed line refers to the black disk mode and the solid one
to the resonant disk mode.  The dotted line (fig. b)
shows the diffractive cross section
with real part enlarged in factor 1.10.
}
   \label{fig8}
   \end{figure*}

Figures \ref{fig8} show the 8 TeV data  \cite{8TeV} at small ${\bf q}^2$. The data are
described in terms of the black disk mode while the resonant disk gives essentially
smaller values of
$\frac{d\sigma}{d{\bf q}^2}$ at $\sim 0.0010$ GeV$^2$.

The data for
$\rho( {\bf q}^2=0,\xi)=A^{H}_{\Re}(0,\xi)/A^{H}_{\Im}(0,\xi)$ prefer
also the  black disk mode:
\be
\begin{tabular}{l|l|l|l}
%\hline
   $\sqrt s$              & 7 TeV      & 8 TeV            &  14 TeV  \\
\hline
$\rho$(data)              &            & 0.12$\pm$ 0.03[\cite{8TeV}] &  \\
$\rho$(black disk)        &  0.17           &  0.17         &  0.16  \\
$\rho$(resonant disk)     &  0.34           & 0.34        &  0.33     \\
%\hline
\end{tabular}
.
\ee
So, the 8 TeV data \cite{8TeV} provides convincing arguents
in favour of the black disk picture
at ultrahigh energies.

The extraction of the real part of the amplitude is based on the
asymptotical relation (2), the question arises how small deviations
from (2) effect the interplay of the hadron and Coulomb interactions.
In Fig. \ref{fig8}b we show the diffractive scattering cross section
with the real part enlarged by a factor 1.10 that demonstrates the
sensitivity of $\frac{d\sigma}{d{\bf q}^2}$ to the value of
$\rho( {\bf q}^2,\xi)$. It also tells that the procedure of the
Eq. (\ref{e2}) determines the real part of the amplitude within
$\sim$10\% accuracy.

\section{Conclusion}

Our studies are performed in terms of the
$K$-matrix function, in hadron physics the $K$-matrix
techniqe was used for non-relativistic processes \cite{ani-ans} and
relativistic ones \cite{aitchison1,aitchison2,ani-sar} as well as for high energy
hadron-hadron collisions \cite{ani-ani} . For ultrahigh energy production processes the
technique was applied in refs.\cite{amn-prod1,amn-prod2} .

The problem of parton cloud structure of hadrons with ultrahigh energy
is a subject of lively discussions
\cite{mart-2014,nemes-2015,gior-megg-2015,dremin-2015,dremin-2015_ufn,tro-tyu-2016} .
A slow growth of the parton cloud radius with energy increase argues
for a glueball origin of the parton clouds
\cite{ann2013DN,ann2014review,a2015ufn} , the glueball origin of the
parton clouds leads to the universality of all hadron total cross
sections \cite{gribov} .

Performing the unitarization of the scattering amplitude
we restrict ourselves to the consideration of the comparatively small momenta
transferred region thus  concentrating our attention on
peripheral interaction of the conventional pomerons.
At large ${\bf q}^2$ other types of the input
pomerons are possible as well as non-pomeron  short-range
contributions (for example, see \cite{KL,KN,GLM,AKL}) but these problems are
beyond the present studies.

We study the asymptotic regime supposing for the parton clouds two
different extreme modes: the black disk mode and resonant disk one. In
the black disk mode the diffractive scattering occurs due to
non-coherent parton interactions while the resonant disk scattering is
the result of coherent interaction of partons with the same impact
parameter $b$; the resonant disk mode realizes the maximal growth of
the amplitude compatible with the Froissart constraint
\cite{Froissart1961} .

In a
number of papers \cite{block,ann2013DN,Dremin2014} the appearance of a
black spot at LHC energy in the profile function at small $b$ is
emphasized but
the pre-LHC and 7 TeV LHC data do not supply us with information to determine
definitely the asymptotic mode of the scattering amplitude.
The 8 Tev data \cite{8TeV} fill a gap in the information noticeably.

The Coulomb and hadronic interaction interplay is realized due to the
interference in the real part of the amplitude, and  the real
part of the resonant disk amplitude is essentially larger than that for
the black disk mode. In the $pp$ scattering the large interference results
in the resonant disk mode as a shoulder in the region ${\bf q}^2\sim 0.0025-0.0075$
GeV$^2$. The shoulder is not observed in 8 TeV data \cite{8TeV}
this is an argument in favour of the realization of the black disk picture
at ultrahigh energies.

\subsubsection*{Acknowledgement}

We thank Y.I. Azimov and A.V. Sarantsev for useful discussions.
The work was supported by RSGSS-4801.2012.2 grant.


\begin{thebibliography}{99}
 \bibitem{8TeV} G. Atchev et al. [TOTEM Collaboration], CERN-PH-EP-2015-325.
\bibitem{Arnison:1983mm}
    G.~Arnison {\it et al.} [UA1 Collaboration],
    %``Elastic and Total Cross-section Measurement at the {CERN} Proton - Anti-proton Collider,''
    Phys.\ Lett.\ B {\bf 128}, 336 (1983).
\bibitem{Bozzo:1984ri}
    M.~Bozzo {\it et al.} [UA4 Collaboration],
    %``Low Momentum Transfer Elastic Scattering at the CERN Proton - anti-Proton Collider,''
    Phys.\ Lett.\ B {\bf 147}, 385 (1984).
\bibitem{Amos:1990fw}
    N.~A.~Amos {\it et al.} [E-710 Collaboration],
    %``$\bar{p}p$ elastic scattering at $\sqrt{s}$ = 1.8-TeV from |t| = $0.034-GeV/c^{2}$ to $0.65-GeV/c^{2}$,''
    Phys.\ Lett.\ B {\bf 247}, 127 (1990).
\bibitem{Augier:1993sz}
    C.~Augier {\it et al.} [UA4/2 Collaboration],
    %``A Precise measurement of the real part of the elastic scattering amplitude at the S anti-p p S,''
    Phys.\ Lett.\ B {\bf 316}, 448 (1993).
\bibitem{Abe:1993xx}
    F.~Abe {\it et al.} [CDF Collaboration],
    %``Measurement of small angle $\bar{p}p$ elastic scattering at $\sqrt{s} = 546$ GeV and 1800 GeV,''
    Phys.\ Rev.\ D {\bf 50}, 5518 (1994).

\bibitem{Latino:2013ued}
     G.~Latino [on behalf of TOTEM Collaboration],
     %``Summary of Physics Results from the TOTEM Experiment,''
     EPJ Web Conf.\  {\bf 49}, 02005 (2013).

\bibitem{Aad:2012pw}
     G.~Aad {\it et al.} [ATLAS Collaboration],
     %``Rapidity gap cross sections measured with the ATLAS detector in $pp$
%collisions at $\sqrt{s}=7$ TeV,''
     Eur.\ Phys.\ J.\ C {\bf 72}, 1926 (2012).

\bibitem{Khachatryan:2015gka}
     V.~Khachatryan {\it et al.} [CMS Collaboration],
%``Measurement of diffraction dissociation cross sections in pp collisions at
%$\sqrt{s}$ = 7 TeV,''
     Phys.\ Rev.\ D {\bf 92},  012003 (2015).

\bibitem{Abelev:2012sea}
     B.~Abelev {\it et al.} [ALICE Collaboration],
     %``Measurement of inelastic, single- and double-diffraction cross sections in
%proton--proton collisions at the LHC with ALICE,''
     Eur.\ Phys.\ J.\ C {\bf 73}, 2456 (2013).

\bibitem{GW}M.L. Good and W.D. Walker, Phys. Rev. {\bf 120}, 1857
(1960).

\bibitem{Dakhno:1999fp}
    L.~G.~Dakhno and V.~A.~Nikonov,
    %``Hadron diffractive processes: The Structure of soft pomeron and color screening,''
    Eur.\ Phys.\ J.\ A {\bf 5}, 209 (1999).

\bibitem{block}M.M. Block and F. Halzen,
Phys. Rev. {\bf D86}, 0501504 (2013).

\bibitem{ann2013asym}
     V.V.~Anisovich, V.A.~Nikonov and J.~Nyiri,
%``Asymptotic regime for hadron-hadron diffractive collisions at ultrahigh energies,''
     Phys.\ Rev.\ D {\bf 88}, 094015 (2013).

\bibitem{Troshin2014}
     S.M.~Troshin and N.E.~Tyurin,
%``Effects of the reflective scattering in hadron production at high energies,''
     Int.\ J.\ Mod.\ Phys.\ A {\bf 29}, 1450151 (2014).

\bibitem{Dremin2014}
     I.M.~Dremin,
%   ``Torus or black disk?,''
     Bull.\ Lebedev Phys.\ Inst.\  {\bf 42}, 21 (2015)
     [Kratk.\ Soobshch.\ Fiz.\  {\bf 42}, 8 (2015)]
     [arXiv:1404.4142 [hep-ph]].

\bibitem{ann2014reso}
     V.V.~Anisovich, V.A.~Nikonov and J.~Nyiri,
%``Hadron collisions at ultrahigh energies: black disk or resonant disk modes?,''
     Phys.\ Rev.\ D {\bf 90}, 074005 (2014).

\bibitem{Gaisser1988}
     T.K.~Gaisser and T.~Stanev,
     %``Mini - Jets in Minimum Bias Events,''
     Phys.\ Lett.\ B {\bf 219}, 375 (1989).

\bibitem{Block1990} M.M.~Block, F.~Halzen and B.~Margolis,
     %``Elastic scattering at s**(1/2) = 1800-GeV: The First look at the
     %asymptotic nucleon,''
     Phys.\ Lett.\ B {\bf 252}, 481 (1990).

\bibitem{Fletcher1992} R.S.~Fletcher,
     %``Diffractive dissociation in the mini - jet model,''
     Phys.\ Rev.\ D {\bf 46}, 187 (1992).

\bibitem{as1978}  V.V.~Anisovich, V.M. Shekhter,
Sov. J. Nucl. Phys. {\bf 28}, 561 (1978); [Yad. Fiz. {\bf 28}, 1079 (1978)].

\bibitem{alr1979}  V.V.~Anisovich, E.M. Levin, M.G. Ryskin,
Sov.J.Nucl.Phys. {\bf 29}, 674 (1979); [Yad.Fiz. {\bf 29}, 1311 (1979)].

\bibitem{ann2013DN}  V.V.~Anisovich, K.V.~Nikonov and V.A.~Nikonov,
%   ``Proton-proton diffractive collisions at ultrahigh energies,''
     Phys.\ Rev.\ D {\bf 88}, 014039 (2013).

\bibitem{ann2014review}
     V.V.~Anisovich, V.~A.~Nikonov and J.~Nyiri,
     Int.\ J.\ Mod.\ Phys.\ A {\bf 29}, 1450096 (2014).

\bibitem{a2015ufn} V.V.~Anisovich, Phys. Usp. {\bf 58}, 963 (2015);
    [UFN {\bf 185}, 1043 (2015)].

\bibitem{Froissart1961}
     M.~Froissart,
     %``Asymptotic behavior and subtractions in the Mandelstam representation,''
     Phys.\ Rev.\  {\bf 123}, 1053 (1961).

\bibitem{ann2015coul-arxiv}
     V.V.~Anisovich, V.~A.~Nikonov and J.~Nyiri,
%``Hadron diffractive scattering at ultrahigh energies, real part
%  of the amplitude and Coulomb interaction'',
     arXiv:1508.02140 [hep-ph].

\bibitem{ann2015real}
     V.V.~Anisovich, V.A.~Nikonov and J.~Nyiri,
%   ``Real part of scattering amplitude at ultrahigh energies,''
     Int.\ J.\ Mod.\ Phys.\ A {\bf 30}, 1550188 (2015).


\bibitem{Anisovich:2016nss}
  V.~V.~Anisovich and V.~A.~Nikonov,
  %``Hadron diffractive scattering at ultrahigh energies and coulomb interaction,''
  arXiv:1601.07015 [hep-ph] (to be published in Mod.\ Phys.\ Lett.\ A).

\bibitem{ann2016arxiv}
  V.~V.~Anisovich, V.~A.~Nikonov and J.~Nyiri,
  %``Asymptotic regimes for diffractive scatterings at ultrahigh energies and coulomb interaction,''
  arXiv:1602.00885 [hep-ph].

\bibitem{Bethe:1958}
     H.A.~Bethe,
     %``Scattering and polarization of protons by nuclei,''
     Annals Phys.\  {\bf 3}, 190 (1958).

\bibitem{Soloviev:1966} L.D. Soloviev,  Zh. Exp. Theor. Fiz. {\bf 49},
292 (1966) [Sov. Phys. JETP {\bf 22}, 205 (1966)].

\bibitem{West:1968}
     G.B.~West and D.~R.~Yennie,
     %``Coulomb interference in high-energy scattering,''
     Phys.\ Rev.\  {\bf 172}, 1413 (1968).

\bibitem{Franco:1973}
     V.~Franco,
     %``Coulomb-nuclear interference,''
     Phys.\ Rev.\ D {\bf 7}, 215 (1973).

\bibitem{Cahn:1982}
     R.~Cahn,
     %``Coulombic - Hadronic Interference in an Eikonal Model,''
     Z.\ Phys.\ C {\bf 15}, 253 (1982).

\bibitem{Kundrat:1993}
     V.~Kundrat and M.~Lokajicek,
     %``High-energy scattering amplitude of unpolarized and charged hadrons,''
     Z.\ Phys.\ C {\bf 63}, 619 (1994).

\bibitem{Kaspar:2011}
     J.~Kaspar, V.~Kundrat, M.~Lokajicek and J.~Prochazka,
%``Phenomenological models of elastic nucleon scattering and predictions for
%LHC,''
     Nucl.\ Phys.\ B {\bf 843}, 84 (2011).

\bibitem{ani-ans} V.V. Anisovich and A.A. Anselm,
Physics-Uspekhi, {\bf 88}, 117 (1966);[UFN {\bf 88},287 (1966)].
\bibitem{aitchison1} I.J.R. Aitchison, Nucl. Phys. {A189}, 417 (1972).
\bibitem{aitchison2} I.J.R. Aitchison, arXiv:1507.02697[hep-ph] (2015).
\bibitem{ani-sar} V.V. Anisovich and A.V. Sarantsev, Eur. Phys. J. {\bf A 16}, 229 (2003).
\bibitem{ani-ani} A.V. Anisovich and V.V. Anisovich
Phys. Lett. {\bf B275}, 491 (1992).

\bibitem{amn-prod1} V.V. Anisovich, M.A. Matveev, V.A. Nikonov Int. J. Mod. Phys.
{\bf A29} 1450176 (2014).
\bibitem{amn-prod2} V.V. Anisovich, M.A. Matveev, V.A. Nikonov Int. J. Mod. Phys.
{\bf A30} 1550054 (2015).

\bibitem{mart-2014}  E. Martynov, arXiv:1412.0269[hep-ph].

\bibitem{nemes-2015}F. Nemes, T. Cs\"org\H{o} and M. Csan\'ad, Int. J.
Mod. Phys. A {\bf 30} 1550076 (2015).

\bibitem{gior-megg-2015} M. Giordano and E. Megiolaro,
Phys. Lett. B {\bf 744}, 263 (2015).

\bibitem{dremin-2015} I.M. Dremin, Adv. High  Energy Phys. 2015, 912743
(2015).

\bibitem{dremin-2015_ufn} I.M. Dremin, Phys. Usp. {\bf 58}, 61 (2015);
[UFN {\bf 185}, 65 (2015)].

\bibitem{tro-tyu-2016} S.M. Troshin and N.E. Tyurin, arXiv:1601/00483 [hep-ph].

\bibitem{gribov} V.N. Gribov, Sov. J. Nucl. Phys. {\bf 17}, 313 (1973).

\bibitem{KL} A.V. Kotikov and L.N. Lipatov, Nucl. Phys.  {\bf B661},
19 (2003).

\bibitem{KN} K. Kang and H. Nastase,  Phys. Lett. {\bf B624}, 125
(2005).

\bibitem{GLM} E. Gotsman, E.M. Levin and U. Maor,
arXiv:1203.2419 (2012).

\bibitem{AKL} A.K. Likhoded, A.V. Luchinsky and A.A. Novoselov,
 Phys. Rev. D{\bf 82}, 114006
(2010).


\end{thebibliography}
\end{document}